\documentclass[a4paper,11pt]{article}
\pdfoutput=1 
\date{}
\usepackage{jheppub}
\usepackage{amsmath,amssymb,amsfonts,amsxtra, mathrsfs,graphics,graphicx,amsthm,epsfig, youngtab,bm,longtable,float,tikz,empheq}
\usepackage[margin=0.5cm]{caption}

\newcommand{\CN}{\mathcal{N}}

\newcommand {\apgt} {\ {\raise-.5ex\hbox{$\buildrel>\over\sim$}}\ }
\newcommand {\aplt} {\ {\raise-.5ex\hbox{$\buildrel<\over\sim$}}\ }

\makeatletter\@addtoreset{equation}{section}\makeatother

\def\Z{\relax\ifmmode\mathchoice
{\hbox{\cmss Z\kern-.4em Z}}{\hbox{\cmss Z\kern-.4em Z}} {\lower.9pt\hbox{\cmsss Z\kern-.4em Z}}
{\lower1.2pt\hbox{\cmsss Z\kern-.4em Z}}\else{\cmss Z\kern-.4em Z}\fi}

\def\e{{\epsilon}}

\def\btimes{~{{{\lower1pt\hbox{$\square$}} \kern-7.6pt \times}}~}

\def\Z{{\Bbb{Z}}}

\def\be{\begin{equation}}
\def\ee{\end{equation}}
\newcommand{\bea}{\begin{eqnarray}}
\newcommand{\eea}{\end{eqnarray}}

\renewcommand{\bar}{\overline}

\renewcommand{\tilde}{\widetilde}

\newcommand{\Reals}{\mathbb{R}}
\newcommand{\Complex}{\mathbb{C}}

\makeatletter
\newcommand{\ellSN}{\mathop{\operator@font sn}\nolimits}
\newcommand{\ellCN}{\mathop{\operator@font cn}\nolimits}
\newcommand{\ellDN}{\mathop{\operator@font dn}\nolimits}
\newcommand{\ellAM}{\mathop{\operator@font am}\nolimits}
\newcommand{\ellK}{\mathop{\smash{\operator@font K}\vphantom{a}}\nolimits}
\newcommand{\ellE}{\mathop{\smash{\operator@font E}\vphantom{a}}\nolimits}
\makeatother

\ifx\genfrac\sdflkaj

\else

\fi

\newcommand{\beq}{\begin{equation}}
\newcommand{\eeq}{\end{equation}}

\makeatletter
\def\mr@ignsp#1 {\ifx\:#1\@empty\else #1\expandafter\mr@ignsp\fi}%
\newcommand{\multiref}[1]{\begingroup
\xdef\mr@no@sparg{\expandafter\mr@ignsp#1 \: }%
\def\mr@comma{}%
\@for\mr@refs:=\mr@no@sparg\do{\mr@comma\def\mr@comma{,}\ref{\mr@refs}}%
\endgroup}
\makeatother

\newcommand{\hypref}[2]{\ifx\href\asklfhas #2\else\href{#1}{#2}\fi}
\newcommand{\Secref}[1]{Section~\multiref{#1}}
\newcommand{\secref}[1]{Sec.~\multiref{#1}}

\newcommand{\appref}[1]{App.~\multiref{#1}}

\newcommand{\figref}[1]{Fig.~\multiref{#1}}
\renewcommand{\eqref}[1]{(\multiref{#1})}

\def\[{\begin{equation}}
\def\]{\end{equation}}
\def\<{\begin{eqnarray}}
\def\>{\end{eqnarray}}

\ifx\href\asklfhas\newcommand{\href}[2]{#2}\fi

\title{Quantum Hydrodynamics from Large-$n$ Supersymmetric Gauge Theories}

\author[a]{Peter Koroteev,}
\author[b]{Antonio Sciarappa}

\affiliation[a]{Perimeter Institute for Theoretical Physics \\ 31 Caroline Street North, Waterloo\\ Ontario N2L2Y5, Canada}
\affiliation[b]{International School of Advanced Studies (SISSA)\\ 265 via Bonomea, Trieste \\ 34136, Italy}

\emailAdd{pkoroteev@perimeterinstitute.ca}
\emailAdd{asciara@sissa.it}

\abstract{We study the connection between periodic finite-difference Intermediate Long Wave ($\Delta$ILW) hydrodynamical systems and integrable many-body models of Calogero and Ruijsenaars-type. The former describe quantum cohomology and quantum K-theory of the ADHM moduli space of Abelian instantons, while the latter arise in the the instanton counting in four and five dimensional supersymmetric gauge theories with eight supercharges in the presence of defects. Using string theory dualities we provide correspondences between hydrodynamical and many-body integrable systems. In particular, we match the energy spectra on both sides.}

\preprint{SISSA  44/2015/FISI}

\begin{document}

\maketitle

\section{Introduction and Summary} \label{Introduction}
In recent years plethora of deep and insightful results was obtained while studying physics of supersymmetric gauge theories in the presence of defects of various types preserving fractional amount of supersymmetry \cite{Gukov:2006jk, Gukov:2008ve, Gaiotto:2013sma, Alday:2010vg, Kanno:2011aa, Bullimore:2015fr}. A typical setup involves a \textit{`higher'} dimensional gauge theory in four, five, or six dimensions with eight supercharges living on a manifold which locally resembles
\begin{equation}
X_{D}=\Reals^4\times \Sigma\,,
\end{equation}
where $\Sigma$ is a manifold of dimension zero, one, or two (usually, a point, a circle, and an elliptic curve respectively). The precise choice of the space-time geometry depends on the problem, in particular on which observable is being computed. The majority of recent work was done on compact manifolds, mostly on spheres \cite{Pestun:2012aa, Hama:2012aa, Kim:2012ab, Kim:2012aa, Imamura:2012aa,Imamura:2012ab,Hosomichi:2012aa,Kallen:2012aa,Kallen:2012ab}.

The second ingredient of the construction is a $d$-dimensional BPS defect which is immersed into the spacetime $X_D$ such that its stress-energy tensor represents a delta-function $T\sim \delta^{(D-d)}T'$. The defect itself supports a \textit{`lower'} dimensional supersymmetric gauge theory on its worldvolume. The defect degrees of freedom interact with the degrees of freedom of the higher dimensional bulk theory, thereby creating a rather complex coupled $D/d$-dimensional system. The gauge interactions in the bulk theory and on the defect are controlled by different couplings, we shall refer to them as $Q$ and $t$ respectively\footnote{An exact expression in terms of gauge coupling constant $\tau_{\text{YM}}$ depends on the number of dimensions.}. Therefore, bulk-defect and bulk-bulk interactions are controlled by $Q$. In the decoupling limit, when $Q\to 0$, those interactions disappear and we are left with the gauge theory on the defect.

For the purpose of this work, in which we shall focus on the five-dimensional $\CN=1^*$ theory with $U(n)$ gauge group, it will be sufficient to study the theory on the following Euclidean space
\begin{equation}
X_5=\Complex_{\epsilon_1}\times \Complex_{\epsilon_2} \times S_\gamma^1\,,
\label{eq:X5flat}
\end{equation}
where we turned on Omega background along two complex directions with equivariant parameters $\epsilon_1$ and $\epsilon_2$ \cite{Nekrasov:2002qd}, and where $\gamma$ is radius of the circle. The 5d theory is enriched by a codimension-two defect which lives in $\Complex_{\epsilon_1} \times S_\gamma^1$. This setup was studied in great details in \cite{Bullimore:2015fr}, where it was used in quantization of the Seiberg-Witten curve of the 5d theory in question as well as finding formal solutions of trigonometric and elliptic quantum many-body systems of Ruijsenaars-Schneider type \cite{MR1329481,MR1322943,MR929148,MR887995,MR851627}.

In this paper we shall capitalize on the results of \cite{Bullimore:2015fr} and study new connections between gauge theories with defects and other physical systems, as well as the interpretation of those connections in mathematical terms. Let us first describe our physics agenda, the mathematical way of stating the new correspondence will be presented later in \secref{Sec:MathResults}.

\subsection{Physics Summary}
One of the natural questions to be asked to a gauge theory is: What happens with the theory when the number of colors becomes large, and what is its effective description? For supersymmetric theories the answer to this question can be formulated purely in physical terms, however, by employing the correspondence with integrable many-body systems \cite{Donagi:1995cf, Gorsky:1995zq} we will also be able to answer this question using integrability language. We will show that as the number of colors (the number of particles in the dual many-body system) becomes large, there is a realization of the 5d $U(n)$ $\CN=1^*$ theory as a certain hydrodynamical model called \textit{periodic finite-difference Intermediate Long Wave model} or $\Delta$ILW. The model is known to be integrable as well \cite{2009JPhA...42N4018S,2009arXiv0911.5005T,2011ntqi.conf..357S}. Dualities between supersymmetric gauge theories with Seiberg-Witten curves \cite{Seiberg:1994aj} and various limits of $\Delta$ILW systems, like Benjamin-Ono limit, have been discussed in the literature, however, the quantum spectrum of the most generic $\Delta$ILW system (and of its differential ILW limit) is not known. In this paper we make a step towards finding the solution of the quantum $\Delta$ILW model using the solution of the elliptic Ruijsenaars-Schneider model presented in \cite{Bullimore:2015fr}.

The integrability of quantum ILW models had been discussed recently in the literature. In \cite{2014JHEP...07..141B,2015arXiv150507116B} the authors discussed the relationship between the $(2,2)$ gauged linear sigma model on $S^2$, whose target space is the ADHM moduli space (ADHM GLSM), and the quantum periodic $N$-dimensional $\text{ILW}_{N}$ system, based on the Bethe/Gauge correspondence \cite{Nekrasov:2009ui,Nekrasov:2009uh}\footnote{The connection between ILW systems and ADHM construction was also mentioned in some recent talks and work in progress \cite{Otalk} and \cite{NOinp,Ntalk}}. Among other things, this correspondence allowed them to compute the spectrum of the $\text{ILW}_{N}$ system in terms of gauge theory quantities; this spectrum is then conjectured to be given by the eigenvalues of $N$ coupled copies of the elliptic Calogero-Sutherland (eCS) system. All the results are obtained as a perturbative series expansion around the known solutions of the Benjamin-Ono and trigonometric Calogero-Sutherland systems. See also \cite{2013JHEP...11..155L,2015JHEP...02..150A} for related results.

In this paper we will see how the results of \cite{2014JHEP...07..141B,2015arXiv150507116B} can be reorganized in a more elegant way, by considering the $\CN=2$ ADHM GLSM on $\Complex \times S^1_{\gamma}$ with $\gamma$ radius of the extra circle, focussing on the one dimensional ILW and $\Delta$ILW systems. The generating function for the ILW spectrum turns out to coincide with the first gauge theory observable $\langle \text{Tr}\,\sigma \rangle$ in three dimensions; moreover, this can be thought as the eigenvalue of the first quantum $\Delta$ILW Hamiltonian $\mathcal{\widehat{H}}_1$ \cite{2009JPhA...42N4018S}, which is therefore expected to be a generating function for the whole set of quantum ILW Hamiltonians $\widehat{I}_l$. 

From the mathematical point of view, the velocity field satisfying the $\Delta$ILW equation and the quantum $\Delta$ILW Hamiltonians $\mathcal{\widehat{H}}_l$ enter in the Fock space representation of the so-called elliptic Ding-Iohara algebra \cite{Ding:1996mq}, whose detailed analysis was performed in \cite{Feigin:2009ab}. This algebra is deeply connected with the free field representation of the elliptic Ruijsenaars-Schneider model. When this system is considered in the limit of the large number of particles, the elliptic Ding-Iohara algebra provides a precise way to relate $\Delta$ILW and elliptic Ruijsenaars-Schneider models at the level of eigenvalues; thereby generalizing and clarifying the connection between the ILW and the elliptic Calogero-Sutherland spectra. 

This connection can be translated in gauge theoretical terms. While the $\Delta$ILW system corresponds to the ADHM quiver on $\Complex \times S^1_{\gamma}$, the $n$-particle eRS system, as we have mentioned earlier, has a gauge theory realization as a 5d $N=1^*$ $U(n)$ theory in Omega background \eqref{eq:X5flat} coupled to a 3d $T[U(n)]$ defect on $\mathbb{C}_{\epsilon_1} \times S^1_{\gamma}$ \cite{Bullimore:2015fr}. One may think of $U(n)$ global symmetry of the 3d theory as being gauged. The eigenfunctions and eigenvalues of the eRS model correspond to the coupled 5d/3d instanton partition function $Z_{5d/3d}^{\text{inst}}$ and to the vacuum expectation values of the Wilson loop in the fundamental representation of $U(n)$ $\langle W^{U(n)}_{\square} \rangle$ respectively, in the so-called Nekrasov-Shatashvili limit \cite{Nekrasov:2009rc} when $\epsilon_2\to 0$. In this work we will show that in the $n \rightarrow \infty$ limit the Wilson loop VEV $\langle W^{U(n)}_{\square} \rangle$ coming from this coupled 5d/3d theory reduces to the $\langle \text{Tr}\,\sigma \rangle$ observable of the twisted chiral ring of the 3d ADHM quiver, thus providing a remarkable connection between these two very different supersymmetric gauge theories.

Line operators $T_k$ act on instanton/vortex partition functions $\mathcal{Z}$ of the 5d/3d theory by quantum shifts of the 3d Fayet-Iliopoulos parameters\footnote{The details will follow in the next section.}
\begin{equation}
T_k \mathcal{Z}= \left\langle W^{U(n)}_k \right\rangle \mathcal{Z}\,,
\label{eq:EigenValEq}
\end{equation}
where $k=1,\dots, n$ is the rank of the antisymmetrization of the fundamental representation of $U(n)$. Thanks to integrability it will be sufficient to look at the fundamental representation. The partition functions are vectors in some (rather large) Hilbert space of states. In order to take the large-$n$ limit of \eqref{eq:EigenValEq}, we need to understand separately large-$n$ behavior of Wilson operator VEVs $\langle W^{U(n)}_{\square} \rangle$ and the states. 

Let us start with the space of states. In the beginning we count (ramified) instantons of the 5d $U(n)$ theory. As we will shortly see, the presence of the $U(1)$ factor in the gauge group will play a crucial role in taking the limit. It will be demonstrated by an explicit calculation in \secref{ThefinitedifferenceILWsystem}, as well as using string theory dualities in \secref{Sec:StringThDer}, that at large $n$ the 5d $U(n)$ theory effectively transforms into a $U(1)$ theory, therefore we expect that the instanton calculus should be reinterpreted accordingly in terms of Abelian noncommutative instantons. One of the noncommutativity parameters will be related to the adjoint mass of the $\CN=1^*$ theory, while the other parameter will be the remaining Omega background velocity $\epsilon_1$. In five dimensions any instanton solution can wrap $S^1_\gamma$ arbitrary many times, so one needs to include the entire Kaluza-Klein tower of those solutions. Given a topological sector $k$ the moduli space of instantons is the Hilbert scheme of $k$ points on $\mathbb{C}^2$ \cite{Nakajima:aa, Schiffmann:2009aa,Schiffmann:2012aa}. The complete moduli space is therefore the union of those Hilbert schemes over all topological sectors. 

The localization formula for a fundamental Wilson loop in the five-dimensional theory in \eqref{eq:X5flat} wrapping $S^1_\gamma$ contains an equivariant character $\chi_{\vec{\lambda}}$ of the universal bundle over the instanton moduli space, which accounts for the propagation of a heavy particle along the circle. 
We expect the expression for character $\chi_{\vec{\lambda}}$ to remain finite after the transition and to depend on the Abelian instanton data. We will be able to prove that as $n\to\infty$ the Wilson loop VEV, up to a certain normalization, becomes
\begin{equation}
\left\langle W^{U(n)}_{\square} \right\rangle\Big\vert_{\lambda} \;\sim\; \mathcal{E}_1^{(\lambda)} = 1 - (1-q)(1-t^{-1}) \sum_s \sigma_s \Big\vert_{\lambda}
\label{eq:MainResultAnn}
\end{equation}
the equivariant Chern character of the universal bundle over the 5d $U(1)$ instanton moduli space. In the formula above $q=e^{i\gamma \epsilon_1}$ and $t=e^{-i\gamma \epsilon}$ with $\epsilon$ being a $\CN=1^*$ mass. In addition Coulomb branch scalars in the 5d theory are set to certain values parameterized by partition $\lambda$. There is a one-to-one correspondence between $\lambda$ and the eigenstate of difference equation \eqref{eq:EigenValEq}. In other words, partition $\lambda$ corresponds to a certain vector in the Fock space, which we have introduced earlier. Finally, $\sigma_s$ in \eqref{eq:MainResultAnn} are Coulomb VEVs of the $U(k)$ gauge group of the $U(1)$ instanton quiver gauge theory\footnote{The ADHM quiver theory is reviewed in \appref{appA}}.

The correspondence between characters of the two 5d theories can be illustrated for the unrefined setup, namely when $q=t$, or, equivalently, when the sum of the first $\Complex^*$ equivariant parameter and the adjoint mass vanishes $\epsilon_1+\epsilon=0$. The $U(n)$ instanton computation involves counting of the fixed points under the action of (complexified) global symmetry $\Complex_{\epsilon_1}^*\times\Complex_{\epsilon}^*\times GL(n,\Complex)$, where we have already removed $\Complex_{\epsilon_2}^*$ action due to the Nekrasov-Shatashvili limit $\epsilon_2\to 0$ which is always implied in equations like \eqref{eq:EigenValEq}. Symmetry $\Complex_{\epsilon_1}^*\times\Complex_{\epsilon}^*$ acts on the ADHM data as $U(1)$ transformations of the corresponding adjoint chiral fields in the ADHM quiver and, as we will see later, for the 5d $U(1)$ ADHM quiver $\epsilon_1$ and $\epsilon$ can be interpreted as twisted masses for those adjoint chirals. The third adjoint chiral, which we call $\chi$, remains massless. 

However, in the refined setup, when $q\neq t$, $\chi$ acquires mass $-\epsilon_1-\epsilon$, thereby breaking the supersymmetry of the 3d ADHM quiver from $\CN=4$ to $\CN=2^*$. This observation will allow us to 
give another interpretation of the right hand side of \eqref{eq:MainResultAnn}. Indeed, the ADHM quiver, as a 3d $\CN=2^*$ theory is self-dual under the thee-dimensional mirror symmetry \cite{Intriligator:1996eu}. Therefore, in order to describe 5d instantons, or Higgs branch of the ADHM quiver, we can study its Coulomb branch and the equations for the supersymmetric vacua of the theory! Thus we will show that $\sigma_s \big\vert_{\lambda}$ from \eqref{eq:MainResultAnn} are obtained as solutions of the (twisted) chiral ring relations. Recall that twisted chiral rings of 3d sigma models on $S^1\times \Reals^2$ describe quantum K-theory of their target manifolds \cite{Nekrasov:2009ui, Gaiotto:2013bwa, Bullimore:2015fr}. We can now arrive to the K-theory version of the results of \cite{2014JHEP...07..141B} stating that the quantum K-theory of the ADHM moduli space is in one to one correspondence with the integrals of motion of the Intermediate Long Wave system.

\subsection{Mathematical Summary}\label{Sec:MathResults}
It is interesting to formulate our results in a more mathematical language. In \cite{Bullimore:2015fr} it was shown that the $n$-particle trigonometric Ruijsenaars-Schneider model computes equivariant quantum K-theory of the cotangent bundle to complete flag variety $T^*\mathbb{F}_n$. In particular, it was demonstrated that level equations for the integrals of motion of the model form a set of relations for a polynomial ring of functions which describe $K_T(T^*\mathbb{F}_n)$, where $T$ is the maximal torus of the global symmetry group $U(n)\times U(1)$. In addition, \cite{Bullimore:2015fr} states that the corresponding K-theoretic Givental $\mathcal{J}$-function is  proportional to the vortex partition function of $U(n)$ 3d $\CN=2^*$ theory, which we discussed above. Therefore it is tempting to conclude that the large-$n$ limit of the quantum K-theory of the cotangent bundle to complete $n$-flag is given by the classical part of the equivariant K-theory of $\widetilde{\mathcal{M}_1}$
\begin{equation}
\lim_{n\to\infty} K_T(T^*\mathbb{F}_n) \simeq K^{\text{cl}}_{q,t}\left(\widetilde{\mathcal{M}_1}\right)\,,
\end{equation}
where
\begin{equation}
\widetilde{\mathcal{M}_1}=\bigoplus\limits_{k=0}^{\infty}\mathcal{M}_{1,k}\,,
\end{equation}
is the direct sum of the moduli spaces of $U(1)$ instantons over all topological sectors.

As for the elliptic Ruijsenaars-Schneider model, to the best of our knowledge, there is no known mathematical object which would describe its spectrum. Let us introduce the following ring\footnote{In \cite{6delliptic} it s suggested that the elliptic Ruijsenaars-Schneider model computes elliptic cohomology of the target manifold.}
\begin{equation}
\mathcal{E}^{Q}_T(T^*\mathbb{F}_n):=\Complex[p^{\pm 1}_i, \tau_i^{\pm1}, Q, t, \mu_i^{\pm 1}]/\mathcal{I}_{\text{eRS}}\,,
\label{eq:eRSRing}
\end{equation} 
where $\mathcal{I}_{\text{eRS}}$ is the ideal generated by the conserved charges of the elliptic Ruijsenaars-Schneider model. The above structure provides a natural elliptic generalization of $K_T(T^*\mathbb{F}_n)$, where $Q$ is the ellipticity parameter. We therefore claim that
\begin{equation}
\lim_{n\to\infty} \mathcal{E}^{Q}_T(T^*\mathbb{F}_n) \simeq K_{q,t}\left(\widetilde{\mathcal{M}_1}\right)\,.
\end{equation}

\subsection{Structure of the Paper}
The rest of the paper is organized as follows. First, in \Secref{Ruijsenaarssystemsfromgaugetheory} we will discuss the trigonometric and elliptic quantum Ruijsenaars-Schneider models constructed using the supersymmetric gauge theory language. Then in \Secref{Freefield realizationofRuijsenaars} we briefly review the basic notions about the trigonometric and elliptic Ding-Iohara algebrae which we will need for our purposes, together with their relation to the Ruijsenaars-Schneider quantum systems. \Secref{ThefinitedifferenceILWsystem} addresses the correspondence between the ADHM theory on $\Complex \times S^1_{\gamma}$, the $\Delta$ILW system and the Ding-Iohara algebra.
Having understood all the ingredients, we shall conclude in \Secref{section5}  by stating, and giving computational evidence for, the proposed correspondence between $\Delta$ILW and eRS in the large number of particles limit. Finally, in \Secref{Sec:Future} we shall list some questions which immediately follow from our results and which will hopefully be addressed in the near future.

\section{Gauge Theory Construction of Ruijsenaars models}\label{Ruijsenaarssystemsfromgaugetheory}
We start with the construction of quantum trigonometric and elliptic Ruijsenaars-Schneider models from supersymmetric gauge theories in three and five dimensions respectively.

\subsection{3d $\CN=2^*$ Theory}
As it was argued in \cite{Bullimore:2015fr} the space of supersymmetric vacua of the 3d $\mathcal{N}=2^*$ $T[U(n)]$ quiver theory on $\Reals^2\times S^1$ describes the phase space of the $n$-particle trigonometric Ruijsenaars-Schneider system.
\begin{figure}[h]
\centering
\includegraphics[width=0.5\textwidth]{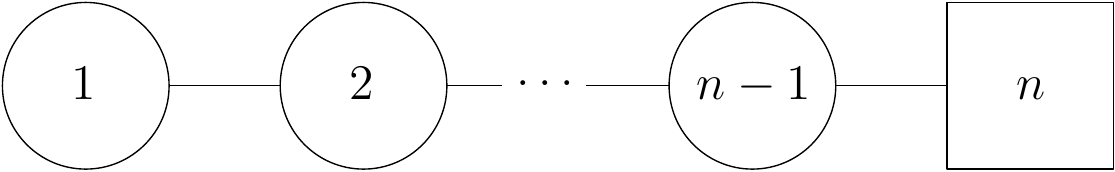}
\caption{The $T[U(n)]$ quiver} 
\end{figure}
The $T[U(n)]$ theory has gauge group $G = \times_{s=1}^{n-1}U(s)$, with an associated $\mathcal{N}=4$ vector multiplet for each factor in $G$, and $\mathcal{N}=4$ hypermultiplets in the bifundamental of $U(s) \times U(s+1)$ with $s=1,\ldots, n-1$, where the last group $U(n)$ is intended as a flavor group. This theory depends on two sets of (exponentiated) parameters: twisted masses $\mu_a$, $a=1,\ldots,n$ for the $U(n)$ flavor group and Fayet-Iliopoulos parameters $\tau_i$ with $i=1\ldots,n$\footnote{Here we introduced an additional topological $U(1)$ as in \cite{Bullimore:2015fr}, so that the physical FI parameter at the $s$-th gauge node is $\tau_{j+1}/\tau_j$.}. In addition, we turn on the canonical $\mathcal{N}=2^*$ deformation parameter $t$, which corresponds to a twisted mass parameter for the adjoint $\mathcal{N}=2$ chiral multiplets contained inside the $\mathcal{N}=4$ vector multiplets\footnote{The bifundamental matter is also charged under the $U(1)_t$ symmetry, see \cite{Gaiotto:2013bwa} for details.}. 

Let us briefly review the connection between the $T[U(n)]$ gauge theory and the trigonometric Ruijsenaars-Schneider system. One needs to analyze the supersymmetric vacua of the $T[U(n)]$ theory on its Coulomb branch. The theory on the Coulomb branch is described by twisted effective superpotential 
\begin{equation}
\widetilde{\mathcal{W}}_{\text{eff}}\left(\mu_s,\,\tau_s,\,t,\,\sigma^{(s)}_a\right)\,, \quad  a=1,\dots, s,\, \quad s=1,\dots, n-1\,,
\end{equation}
where $\sigma^{(s)}_a$ are scalars in the vector multiplets of the Cartan subalgebra of $G$. It was shown in \cite{Bullimore:2015fr} that equations 
\begin{equation}
\text{exp} \left( \sigma^{(s)}_{\alpha}\dfrac{\partial \widetilde{\mathcal{W}}_{\text{eff}}}{\partial \sigma^{(s)}_{\alpha}} \right) = 1\,,
\end{equation}
determining the supersymmetric vacua, i.e. the twisted chiral ring relations, reduce to classical Hamiltonian equations of the trigonometric Ruijsenaars-Schneider model
\begin{equation}
D_{n,\vec{\tau}}^{(k)}=S_k(\mu_1,\dots, \mu_n)\,, \quad k=1,\dots, n\,,
\end{equation}
where $S_k$ are symmetric polynomials of degree $k$ of its variables, for example $S_k(\mu_1,\dots, \mu_n)=\mu_1+\dots+ \mu_n$, and the left hand side presents $n$ integrals of motion of the trigonometric Ruijsenaars-Schneider model. The first Hamiltonian reads
\begin{equation}
D_{n}^{(1)}(\tau_i, p_\tau^i) = \sum_{i=1}^n \prod_{j \neq i}^n \dfrac{t \tau_i - \tau_j}{\tau_i - \tau_j} p^i_{\tau}\,,
\end{equation}
and depends on coordinates $\tau_i$ and momenta $p^i_{\tau}$. The latter are given as exponentiated derivatives of the on-shell twisted superpotential
\begin{equation}
p^i_{\tau} = \text{exp} \left( \tau_i\dfrac{\partial \widetilde{\mathcal{W}}_{\text{eff}}}{\partial \tau_i} \right)\,.
\end{equation}

In order to quantize the integrable system we study the $T[U(n)]$ theory in the Omega background $\Complex_q\times S^1$ \cite{Nekrasov:2009ui, Bullimore:2015fr}. We replace momenta $p_{\tau_i}$ by quantum shift operators $T_{q,i}$, with $q \sim e^{i \gamma \widetilde{\epsilon}_1}$; the vacua equations now become operator equations annihilating the partition function of the $T[U(n)]$ theory. In other words, the $T[U(n)]$ partition function is the common eigenfunction of the quantum trigonometric Ruijsenaars-Schneider Hamiltonians
\begin{equation}
\label{tRSh}
D_{n,\vec{\tau}}^{(1)}(q,t) = \sum_{i=1}^n \prod_{j \neq i}^n \dfrac{t \tau_i - \tau_j}{\tau_i - \tau_j} T_{q,i}\,.
\end{equation}
Here $\tau_l$ are the positions of the particles, $t$ is a parameter determining the strength of the interaction, and $T_{q,i}$ is a shift operator acting as 
\begin{equation}
T_{q,i} f(\tau_1, \ldots, \tau_i, \ldots, \tau_n) = f(\tau_1, \ldots, q\tau_i, \ldots, \tau_n)
\end{equation}
on functions of the $\tau_l$ variables; we can think of it as $T_{q,i} = e^{i \gamma \widetilde{\epsilon}_1 \, \tau_i \partial_{\tau_i}} = q^{ \tau_i \partial_{\tau_i}}$. In the limit $\gamma \rightarrow 0$, $D_{n,\vec{\tau}}^{(1)}$ reduces to the $n$-particles trigonometric Calogero-Sutherland Hamiltonian. The eigenvalue of quantum Hamiltonian $D_{n}^{(1)}$ is given by the vacuum expectation value $\left\langle W_{\square}^{U(n)} \right\rangle = S_1(\mu_i)=\mu_1 + \ldots + \mu_n$ of the flavor Wilson loop wrapping $S^1_{\gamma}$ in the fundamental representation of $U(n)$. 

The partition function on $\Complex_q\times S^1$ coincides (up to a pre-factor) with the holomorphic blocks $B_{l}$ \cite{Beem:2012uq} of the $T[U(n)]$ theory ($l=1\,\ldots, n!$), which in turn can be obtained by factorizing the partition function on the squashed three-sphere $S^3_b$ as
\begin{equation}
\mathcal{Z}_{S^3_b}(\vec{\mu},\vec{\tau},t,q) = \sum_{l=1}^{n!} \left\vert \mathcal{B}_l (\vec{\mu},\vec{\tau},t,q) \right\vert^2
\end{equation}  
after an appropriate identification of $\widetilde{\epsilon}_1$ with the squashing parameter $b$. Note that the holomorphic blocks $B_{l}$ are infinite series in the FI parameters and have to be thought as formal eigenstates, as they might not be normalizable. However, when the mass parameters are specified to certain values the above series expansion truncates to Macdonald polynomials.

\subsection{Macdonald Polynomials}
The holomorphic block for $T[U(2)]$ theory with FI parameter $\tau_1/\tau_2$ and mass parameters $\mu_1,\mu_2, t$ on $\mathbb{C}_q\times S^1$ reads \cite{Bullimore:2015fr}
\begin{equation}
\label{eq:HolBlockTU2}
\mathcal{B}(\tau_1,\tau_2;\mu_1,\mu_2,t,q)= \frac{\Theta_q(t^{-1/2} \, \tau_1)\Theta_q(t^{1/2} \, \tau_2) }{\Theta_q(\mu_2\tau_1)\Theta_q(\mu_1\tau_2)} {}_2F_1\left(t,t\frac{\mu_2}{\mu_1};q\frac{\mu_2}{\mu_1};q;qt^{-1} \frac{\tau_1}{\tau_2}\right)\,,
\end{equation}
where
\begin{equation}
\Theta_q(x) = (q;q)_{\infty} (x;q)_{\infty} (q/x;q)_{\infty}\,, \quad (x;q)_{\infty} = \prod_{s=0}^{\infty} (1-x q^s)\,.
\end{equation}
The second holomorphic block is obtained from the above expression by interchanging $\mu_1$ and $\mu_2$. Both blocks satisfy difference equations of trigonometric Ruijsenaars-Schneider system
\begin{align}
D^{(1)}_{q} \mathcal{B} &= (\mu_1+\mu_2)\mathcal{B}\,,\cr
D^{(2)}_{q} \mathcal{B} &= \mu_1\mu_2\mathcal{B}\,,
\end{align}
where $D^{(1,2)}_{q}$ are tRS Hamiltonians, they commute between each other. For completeness, let us mention that the operator \eqref{tRSh} is the first of a set of $n$ commuting operators, given by 
\begin{equation}
D_{n,\vec{\tau}}^{(r)}(q,t) = t^{r(r-1)/2} \sum_{\substack{I \subset \{1,2,\ldots, n\} \\ \# I = r}} \prod_{\substack{ i \in I \\ j \notin I}} \dfrac{t \tau_i - \tau_j}{\tau_i - \tau_j} \prod_{i \in I} T_{q,i} \;\;\; \text{for} \;\;\; r = 1, \ldots, n \label{tRShs}
\end{equation}
In mathematical literature, the operator $D_{n,\vec{\tau}}^{(1)}$ is known as the first Macdonald difference operator; its eigenfunctions, known as Macdonald polynomials, are given by symmetric polynomials in $n$ variables $\tau_l$ of total degree $k \leqslant n$, and are in one-to-one correspondence with partitions $\lambda = (\lambda_1,\ldots,\lambda_n)$ of $k$ of length $n$. 

Now we can make the following observation\footnote{See the end of Section 3 of \cite{Bullimore:2015fr}.}. For a given partition $\lambda$ we identify parameters $\mu_a$ as follows
\begin{equation}
\label{mu}
\mu_a = q^{\lambda_a} t^{n-a} \;\;\;,\;\;\; a = 1, \ldots, n \,.
\end{equation}
Having done so we see that the series expansion of holomorphic block \eqref{eq:HolBlockTU2} truncates and it turns into a Macdonald polynomial
 $P_{\lambda}(\vec{\tau};q,t)$ corresponding to the partition $\lambda$ as
\begin{equation}
D_{n,\vec{\tau}}^{(1)}(q,t) P_{\lambda}(\vec{\tau}; q, t) = E_{tRS}^{(\lambda; n)} P_{\lambda}(\vec{\tau}; q, t) \label{tRSeq}
\end{equation}
with an eigenvalue given by
\begin{equation}
E_{tRS}^{(\lambda; n)} = \sum_{j=1}^n q^{\lambda_j} t^{n-j} \label{tRSev}
\end{equation}
Thus for $k=2$ we get 
\begin{align}
\label{eq:MacHolBlock}
\mathcal{B}(\tau_1,\tau_2;t^{-1/2}q,t^{1/2}q)&=P_{\tiny\yng(2)}(\tau_1,\tau_2; q,t)\,,\cr
\mathcal{B}(\tau_1,\tau_2;t^{-1/2},t^{-1/2}q^2)&=P_{\tiny\yng(1,1)}(\tau_1,\tau_2\vert q,t)\,.
\end{align}
In what follows it is instructive make the following change of variables
\begin{equation}
p_m = \sum_{l=1}^n \tau_l^m\,,
\label{eq:PowerSum}
\end{equation} 
For $k=2$ we have two partitions ${\tiny\yng(2)}$ and ${\tiny\yng(1,1)}$, corresponding to the Macdonald polynomials in \eqref{eq:MacHolBlock}
\begin{equation}
\label{mac}
P_{\tiny\yng(2)}=\dfrac{1}{2}(p_1^2-p_2)\,, \qquad
P_{\tiny\yng(1,1)}=\dfrac{1}{2}(p_1^2-p_2) + \dfrac{1-qt}{(1+q)(1-t)}p_2\,. 
\end{equation} 
Most importantly, this expression in terms of power sum symmetric polynomials \eqref{eq:PowerSum} is the same for any $n$. 

Below we list several examples for degree $k=2$ Macdonald polynomials for $n=2$ and $n=3$
\begin{itemize}
\item For $n=2$ the eigenfunction for the partition $(1,1)$ and its eigenvalues are
\begin{equation}
P_{(1,1)}(\tau_1, \tau_2; q, t) = \tau_1 \tau_2 \;\;\;,\;\;\; E_{tRS}^{((1,1); 2)} = qt + q
\end{equation}
while for the partition $(2,0)$ we have
\begin{equation}
P_{(2,0)}(\tau_1, \tau_2; q, t) = \tau_1 \tau_2 + \dfrac{1-qt}{(1+q)(1-t)} (\tau_1^2 + \tau_2^2) \;\;\;,\;\;\; E_{tRS}^{((2,0); 2)} = q^2t + 1
\end{equation}
\item For $n=3$ the partition $(1,1,0)$ has eigenfunction
\begin{equation}
P_{(1,1,0)}(\tau_1, \tau_2, \tau_3; q, t) = \tau_1 \tau_2 + \tau_1 \tau_3 + \tau_2 \tau_3 
\end{equation}
and eigenvalue
\begin{equation}
E_{tRS}^{((1,1,0); 2)} = qt^2 + q t + 1
\end{equation}
while the partition $(2,0,0)$ has eigenfunction
\begin{equation}
P_{(2,0,0)}(\tau_1, \tau_2, \tau_3; q, t) = \tau_1 \tau_2 + \tau_1 \tau_3 + \tau_2 \tau_3 + \dfrac{1-qt}{(1+q)(1-t)} (\tau_1^2 + \tau_2^2 + \tau_3^2)
\end{equation}
and eigenvalue
\begin{equation}
E_{tRS}^{((2,0,0); 2)} = q^2t^2 + t + 1\,.
\end{equation}
\end{itemize}
The generic case follows along these lines. 

To conclude, the tRS/gauge theory dictionary can be summarized as follows:
\begin{center}
\renewcommand\arraystretch{1.2}
\begin{tabular}{|c|c|}
\hline
\textbf{quantum tRS model} & \textbf{3d $\CN=2^*$ $T[U(n)]$ theory} \\ 
\hline number of particles $n$ & rank 3d flavor group \\ 
\hline particle positions $\tau_j$ & 3d Fayet-Iliopoulos parameters \\
\hline interaction coupling $t$ & 3d $\mathcal{N}=2^*$ deformation parameter \\
\hline shift parameter $q$ & Omega background $e^{i \gamma \widetilde{\epsilon}_1}$ \\
\hline eigenvalue $E_{tRS}^{(\lambda; n)}$ & $\langle W_{\square}^{U(n)} \rangle$ for flavour $U(n)$ at fixed $\mu_a$ \\
\hline 
eigenfunctions $P_{\lambda}(\vec{\tau}; q, t)$ & holomorphic blocks $B_{l}$ at fixed $\mu_a$\\
\hline
\end{tabular}
\renewcommand\arraystretch{1}
\end{center}

\subsection{Elliptic Generalization}
\begin{figure}[h]
\centering
\includegraphics[width=0.9\textwidth]{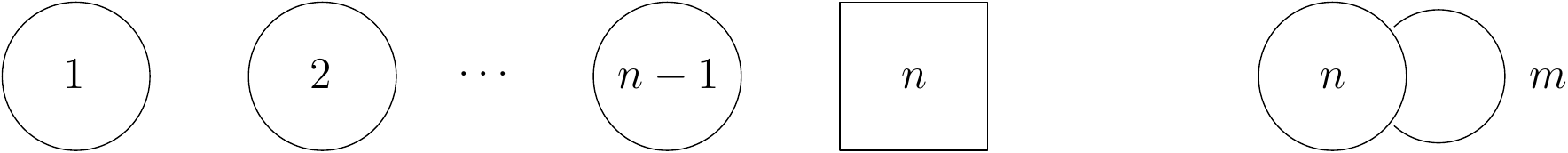}
\caption{The 3d $T[U(n)]$ theory as a defect for the 5d $U(n)$ $\mathcal{N}=1^*$ theory} 
\label{fig:Coupled5d3d}
\end{figure}

As we have mentioned in the introduction, quantum spectrum for the elliptic Ruijsenaars-Schneider model can be computed by studying the 5d $\mathcal{N}=1^*$ $U(n)$ theory on $\mathbb{C}^2_{\widetilde{\epsilon}_1,\widetilde{\epsilon}_2} \times S^1_{\gamma}$ in the Nekrasov-Shatashvili imit $\widetilde{\epsilon}_2 \rightarrow 0$ in presence of codimension-two defect. When the 5d gauge interactions are turned off, the theory reduces on the defect and we are left with the 3d $\CN=2^*$ $T[U(n)]$ theory, which we have discussed above in details. The 5d/3d system can be also thought of as both theories coupled together by gauging the $U(n)$ flavor symmetry of $T[U(n)]$ \figref{fig:Coupled5d3d}. The mass $m$ for the adjoint field in the 5d $\mathcal{N}=2$ vector multiplet breaks supersymmetry from $\mathcal{N}=2$ to $\mathcal{N}=1^*$ and coincides with the parameter $t$ of the 3d $\mathcal{N}=2^*$ deformation as $t \sim e^{-i \gamma m}$, while the 3d twisted masses $\mu_a$ represent VEVs of to the 5d Coulomb branch moduli. 
In the coupled system, which reproduces the eRS model, the elliptic deformation parameter $p$ is given by $Q = e^{-8 \pi^2 \gamma /g_{YM}^2}$ with $g_{YM}$ being the 5d Yang-Mills coupling. 

The elliptic Ruijsenaars-Schneider Hamiltonians are elliptic generalizations of \eqref{tRSh} and are defined as 
\begin{equation}
\label{eRSh}
D_{n,\vec{\tau}}^{(1)}(q,t;p) = \sum_{i=1}^n \prod_{j \neq i}^n \dfrac{\Theta_p(t \tau_i/\tau_j)}{\Theta_p(\tau_i/\tau_j)} T_{q,i} 
\end{equation}
For $p=0$, the Hamiltonian \eqref{eRSh} reduces to \eqref{tRSh}. The solution to this model, i.e. eigenfunctions and eigenvalues of \eqref{eRSh}, at the present stage can only be obtained perturbatively around the known tRS solution by expanding \eqref{eRSh} around $p \sim 0$ \cite{Bullimore:2015fr}. It turns out that eigenfunctions, evaluated at locus \eqref{mu}, can still be labelled by partitions of $k$ of length $n$, although this time the eigenfunctions are symmetric polynomials in the $\tau_l/\tau_m$ variables. We shall discuss this solution momentarily.

Interestingly \eqref{mu} has its own meaning in the 5d theory -- the equations provide the condition for Higgs branch of the theory. Indeed, as it was discussed in \cite{Chen:2012we} using brane constructions, when 5d Coulomb branch parameters are set to the values given in \eqref{mu}, Higgs branches open up and the theory admits vortex solutions. We shall discuss the implications of this fact and present further string theoretic derivation of our results in \secref{Sec:StringThDer}.

The eigenfunctions and eigenvalues of \eqref{eRSh} were computed in \cite{Bullimore:2015fr} perturbatively in $Q$. 
The former are instanton partition functions of the coupled 5d/3d system, while the latter are VEVs of Wilson loops wrapping $S^1_\gamma$.
The first correction in $Q$ to the eigenvalue 
\begin{equation}
\left\langle W_{\square}^{SU(n)} \right\rangle  = \left\langle W_{\square}^{U(n)} \right\rangle \Big/ \left\langle W_{\square}^{U(1)} \right\rangle \label{wilson}
\end{equation}
is obtained from
\begin{equation}
\left\langle W_{\square}^{U(n)} \right\rangle = \sum_{a=1}^n \mu_a - Q \dfrac{(q-t)(1-t)}{q t^n} \sum_{a=1}^n \mu_a
\prod_{\substack{b=1 \\ b \neq a}}^n \dfrac{(\mu_a - t \mu_b)(t \mu_a - q \mu_b)}{(\mu_a - \mu_b)(\mu_a - q \mu_b)} + o(Q^2)\,,
\end{equation}
\begin{equation}
\left\langle W_{\square}^{U(1)} \right\rangle = \dfrac{(Qt^{-1};Q)_{\infty}(Qtq^{-1};Q)_{\infty}}{(Q;Q)_{\infty}(Qq^{-1};Q)_{\infty}}\,.
\end{equation}
This formula will become important later in this chapter. 

In summary, the eRS/gauge theory dictionary reads as follows
\begin{center}
\renewcommand\arraystretch{1.2}
\begin{tabular}{|c|c|}
\hline
\textbf{quantum eRS model} & \textbf{5d/3d theory} \\ 
\hline number of particles $n$ & rank 3d flavor group / 5d gauge group  \\ 
\hline particle positions $\tau_j$ & 3d Fayet-Iliopoulos parameters \\
\hline interaction coupling $t$ & 3d $\mathcal{N}=2^*$ / 5d $\mathcal{N}=1^*$ deformation $e^{- i \gamma m}$ \\
\hline shift parameter $q$ & Omega background $e^{i \gamma \widetilde{\epsilon}_1}$ \\
\hline elliptic deformation $p$ & 5d instanton parameter $Q = e^{-8\pi^2 \gamma/g^2_{YM}}$ \\
\hline eigenvalues $E_{tRS}^{(\lambda; n)}$ & $\langle W_{\square}^{U(n)} \rangle$ for 5d $U(n)$ in NS limit at fixed $\mu_a$ \\
\hline eigenfunctions & $Z^{5d/3d}_{\text{inst}}$ in NS limit at fixed $\mu_a$\\
\hline
\end{tabular}
\renewcommand\arraystretch{1}
\end{center} 

\section{Collective Coordinate Realization of Ruijsenaars Systems}\label{FreefieldrealizationofRuijsenaars}
In the previous section we discussed in some detail the $n$-particle quantum trigonometric and elliptic Ruijsenaars-Schneider models from both the integrable system and gauge theory point of view. As reviewed there, the gauge theoretic reformulation allows us to explicitly compute eigenfunctions and eigenvalues of the elliptic Ruijsenaars model, perturbatively in the elliptic parameter around the trigonometric solution, thanks to our good understanding of instanton computations in supersymmetric gauge theories. 

In this section we will consider these systems in the limit in which the number of particles is sent to infinity. This is in order to make contact with quantum integrable systems of hydrodynamic type, in particular with the quantum Intermediate Long Wave system (ILW) and its finite-difference version ($\Delta$ILW): these will be described in \secref{ThefinitedifferenceILWsystem}. 
In fact, at the classical level it is known that the dynamics of the classical trigonometric Calogero-Sutherland model in the $n \rightarrow \infty$ limit is equivalent to the classical Benjamin-Ono (BO) equation (a particular limit of ILW) \cite{2009JPhA...42m5201A}; similarly, the large $n$ dynamics of classical elliptic Calogero is given by the classical ILW equation \cite{2014JHEP...07..141B}. Although there are no computations in the literature to the best of our knowledge (especially because of the little attention received by the $\Delta$ILW system), if one thinks of the trigonometric and elliptic Ruijsenaars-Schneider models as finite-difference versions of Calogero-Sutherland, one can expect similar classical large $n$ relations to hold between trigonometric/elliptic Ruijsenaars and $\Delta$BO/$\Delta$ILW systems. 

At the quantum level, the appropriate formalism to study the $n \rightarrow \infty$ limit of Calogero is the \textit{collective field theory} (or \textit{bosonization}) approach \cite{Jevicki:1979mb,Minahan:1994hn,1995PhLB..352..111I}. The essence of this formalism consists in solving the quantum system by regarding the eigenfunctions as functions of all possible symmetric combinations of the coordinates; it is then easy to consider the large $n$ limit in terms of this basis of symmetric functions. The quantum trigonometric Calogero system has been analysed with the collective field method in \cite{1995PhLB..347...49A}; although not explicitly written there, it is easy to recognize that the trigonometric Calogero Hamiltonian written in collective coordinates coincides with the second conserved quantity of quantum BO. The more complicated case of quantum elliptic Calogero has been partially analysed in \cite{2000math.ph...7036L,2004CMaPh.247..321L}. \\
The collective coordinate description of quantum trigonometric and elliptic Ruijsenaars-Schneider models has been discussed in mathematical terms in \cite{Feigin:2009ab}. This is given in terms of a deformed Heisenberg algebra, and is found to be deeply related to a particular representation of the so-called Ding-Iohara algebra (trigonometric and elliptic). Our representation of the trigonometric algebra allows one to consider trigonometric Ruijsenaars at $n \rightarrow \infty$, and in this limit it reduces to a quantum integrable system with an infinite number of commuting Hamiltonians, which has later been interpreted as the finite-difference BO system in \cite{2009arXiv0911.5005T,2011ntqi.conf..357S}. Similarly, the twisted elliptic deformation of the Ding-Iohara algebra has been recognized as the finite-difference version of ILW in \cite{2009JPhA...42N4018S}. 

Here we will briefly review the results of \cite{Feigin:2009ab} which are relevant for our discussion; the finite-difference versions of BO and ILW will be introduced in the next section. 

\subsection{The Trigonometric Ruijsenaars System} 
\label{The trigonometric case}
Let us start by considering the collective coordinate description of tRS. In order to do so, we will first need to introduce the Macdonald symmetric functions; we will follow the conventions of \cite{2013arXiv1301.4912S,2014SIGMA..10..021S,2013arXiv1309.7094S}. Let 
\begin{equation}
\Lambda_n (q,t) = \mathbb{Q}(q,t)[\tau_1, \ldots, \tau_n]^{\mathfrak{S}_n}
\end{equation}
be the space symmetric polynomials over $\mathbb{Q}(q,t)$ of $n$-variables with $\mathfrak{S}_n$ the $n$-th symmetric group. As in \secref{Ruijsenaarssystemsfromgaugetheory}, let us introduce the power sum symmetric polynomials 
\begin{equation}
p_m = \sum_{l=1}^n \tau_l^m\,,
\end{equation}
and define $p_{\lambda} = p_{\lambda_1} \cdots p_{\lambda_{n}}$ for a partition of size $\vert \lambda \vert = k$ and length $n$. \\
Now, let $\rho_n^{n+1}: \Lambda_{n+1}(q,t) \rightarrow \Lambda_n(q,t)$ be the homomorphism given by
\begin{equation}
(\rho_n^{n+1}f)(\tau_1, \ldots, \tau_n) = f(\tau_1, \ldots, \tau_n, 0) \;\;\; \text{for} \;\;\; f \in \Lambda_{n+1}(q,t)\,,
\end{equation}
and define the ring of symmetric functions $\Lambda(q,t)$ as the projective limit defined by $\{ \rho_n^{n+1} \}_{n\geqslant 1}$
\begin{equation}
\Lambda(q,t) = \lim_{\longleftarrow n} \Lambda_n (q,t)
\end{equation}
Set $\{ p_{\lambda} \}$ forms a basis of $\Lambda(q,t)$. By defining $n_{\lambda}(a) = \# \{ i:\lambda_i = a \}$ and
\begin{equation}
z_{\lambda} = \prod_{a \geqslant 1} a^{n_{\lambda}(a)}n_{\lambda}(a)!\,, \;\;\;\;\;\; z_{\lambda}(q,t) = z_{\lambda} \prod_{i=1}^{l(\lambda)} \dfrac{1-q^{\lambda_i}}{1-t^{\lambda_i}}\,,
\end{equation} 
we can introduce the inner product
\begin{equation}
\langle p_{\lambda}, p_{\mu} \rangle_{q,t} = \delta_{\lambda, \mu} z_{\lambda}(q,t)\,. \label{inpro}
\end{equation}
Set $\{ \tilde{p}_{\lambda} \} = \{ z_{\lambda}^{-1}(q,t) p_{\lambda} \}$ will therefore be a dual basis with respect to $\{ p_{\lambda} \}$ under the inner product \eqref{inpro}; moreover we have
\begin{equation}
\sum_{\lambda} p_{\lambda}(\tau) \tilde{p}_{\lambda}(\tilde{\tau}) = \textstyle{\prod} (q,t)(\tau,\tilde{\tau})
\end{equation}
in terms of the so-called reproduction kernel
\begin{equation}
{\textstyle{\prod}} (q,t)(\tau,\tilde{\tau}) = \prod_{i,j \geqslant 1} \dfrac{(t \tau_i \tilde{\tau}_j;q)_{\infty}}{(\tau_i \tilde{\tau}_j;q)_{\infty}}\,, \;\;\;\;\;\; (a;q)_{\infty} = \prod_{s \geqslant 0} (1-aq^s)\,.
\end{equation}
The statement holds in general: given two bases $\{ u_{\lambda} \}$, $\{ v_{\lambda} \}$ of $\Lambda(q,t)$, they are dual under \eqref{inpro} if and only if $\sum_{\lambda} u_{\lambda}(\tau) v_{\lambda}(\tilde{\tau}) = \textstyle{\prod} (q,t)(\tau,\tilde{\tau})$; in this sense, the form of the inner product is determined by the form of the kernel function. For our discussion, the most relevant basis of symmetric functions is given by the Macdonald basis $\{ P_{\lambda}(\tau; q,t) \}$, uniquely determined by the following conditions
\begin{equation}
\begin{split}
(1) & \;\;\; P_{\lambda}(\tau; q,t) = m_{\lambda}(\tau) + \sum_{\mu < \lambda} u_{\lambda \mu}(q,t) m_{\mu}(\tau) \;\;\; \text{with} \;\;\; u_{\lambda \mu}(q,t) \in \mathbb{Q}(q,t)\,, \\
(2) & \;\;\; \langle P_{\lambda}(\tau; q,t), P_{\mu}(\tau; q,t) \rangle_{q,t} = 0 \;\;\; \text{for} \;\;\; \lambda \neq \mu\,,
\end{split}
\end{equation}
where $m_{\lambda}(\tau)$ are monomial symmetric functions and $\lambda > \mu$ $\Longleftrightarrow$ $\vert \lambda \vert = \vert \mu \vert$ with $\lambda_1 + \ldots + \lambda_i \geqslant \mu_1 + \ldots + \mu_i$ for all $i$. From the functions $P_{\lambda}(\tau;q,t)$ we recover the $n$-variables Macdonald polynomials as $P_{\lambda}(\tau_1, \ldots, \tau_n; q,t) = P_{\lambda}(\tau_1, \ldots, \tau_n, 0, 0, \ldots; q,t)$; these are eigenstates of the Hamiltonians \eqref{tRSh}, \eqref{tRShs} and satisfy \eqref{tRSeq}. 

\subsubsection{Free Field Realization}
We are now ready to discuss the collective coordinate (or free boson) realization of the tRS Hamiltonian \eqref{tRSh}. The idea here is to introduce a $(q,t)$-deformed version of the Heisenberg algebra $\mathcal{H}(q,t)$, with generators $a_m$ ($m \in \mathbb{Z}$) and commutation relations
\begin{equation}
[a_m,a_n] = m \dfrac{1-q^{\vert m \vert}}{1-t^{\vert m \vert}} \delta_{m+n,0}\,. \label{defHei}
\end{equation}
A canonical basis in the Fock space of $\mathcal{H}(q,t)$ is given by the set of states $a_{-\lambda} \vert 0 \rangle = a_{-\lambda_1} \cdots a_{-\lambda_{l(\lambda)}} \vert 0 \rangle$ depending on a partition $\lambda$; a generic state will be a linear combination of the basis ones, with coefficients in $\mathbb{Q}(q,t)$. 
Let us notice that the bra-ket product among basis states is such that
\begin{equation}
\langle 0 \vert 0 \rangle = 1\,, \;\;\;\;\;\; \langle 0 \vert a_{\lambda} a_{-\mu} \vert 0 \rangle = \delta_{\lambda, \mu} z_{\lambda}(q,t)\,,
\end{equation} 
and therefore coincides with the inner product \eqref{inpro}. This is in agreement with the natural isomorphism between this Fock space and $\Lambda(q,t)$, simply given by
\begin{equation}
a_{-\lambda} \vert 0 \rangle \; \longleftrightarrow \; p_{\lambda} \label{iso}
\end{equation}
for fixed partition $\lambda$. Now, in order to reproduce the action of $D^{(1)}_{n,\vec{\tau}}$ in terms of bosonic operators, we follow \cite{Feigin:2009ab} (see also \cite{2013arXiv1301.4912S,2014SIGMA..10..021S,2013arXiv1309.7094S}) and introduce the vertex operators
\begin{equation}
\begin{split}
\eta(z) & \,=\, \text{exp} \left( \sum_{n>0} \dfrac{1-t^{-n}}{n} a_{-n}z^n \right) \text{exp} \left( - \sum_{n>0} \dfrac{1-t^n}{n} a_n z^{-n} \right) \\
& \,=\, :\text{exp}\left( - \sum_{n \neq 0} \dfrac{1-t^n}{n} a_n z^{-n} \right): \,=\, \sum_{n \in \mathbb{Z}} \eta_{n}z^{-n}
\end{split}
\end{equation}
and
\begin{equation}
\begin{split}
\xi(z) & \,=\, \text{exp} \left( -\sum_{n>0} \dfrac{1-t^{-n}}{n} (tq^{-1})^{n/2} a_{-n}z^n \right) \text{exp} \left(  \sum_{n>0} \dfrac{1-t^n}{n} (tq^{-1})^{n/2} a_n z^{-n} \right) \\
& \,=\, :\text{exp}\left( \sum_{n \neq 0} \dfrac{1-t^n}{n} (tq^{-1})^{\vert n \vert /2} a_n z^{-n} \right): \,=\, \sum_{n \in \mathbb{Z}} \xi_{n}z^{-n}\,,
\end{split}
\end{equation}
together with
\begin{equation}
\phi(z) = \text{exp} \left( \sum_{n>0} \dfrac{1-t^n}{1-q^n}a_{-n}\dfrac{z^n}{n} \right)\,, \;\;\;\;\;\; 
\phi^*(z) = \text{exp} \left( \sum_{n>0} \dfrac{1-t^n}{1-q^n}a_{n}\dfrac{z^n}{n} \right)\,. \label{phi}
\end{equation}
By defining $\phi_n(\tau) = \prod_{i=1}^n \phi(\tau_i)$ one can show that the kernel function is reproduced by operators $\phi_n(\tau)$, $\phi_n^*(\tau)$ as 
\begin{equation}
\langle 0 \vert \phi_n^*(\tau) \phi_n(\tilde{\tau}) \vert 0 \rangle = {\textstyle{\prod}} (q,t)(\tau,\tilde{\tau})\,,
\end{equation}  
while the action of $D^{(1)}_{n,\vec{\tau}}$ in terms of oscillators can be expressed by the formulae
\begin{equation}
\begin{split}
& [\eta(z)]_1 \phi_n(\tau) \vert 0 \rangle = \left[ t^{-n} + t^{-n+1}(1-t^{-1})D^{(1)}_{n,\vec{\tau}}(q,t) \right] \phi_n(\tau) \vert 0 \rangle\,, \\
& [\xi(z)]_1 \phi_n(\tau) \vert 0 \rangle = \left[ t^{n} + t^{n-1}(1-t)D^{(1)}_{n,\vec{\tau}}(q^{-1},t^{-1}) \right] \phi_n(\tau) \vert 0 \rangle\,, \label{keytr}
\end{split}
\end{equation}
where $[\;\;]_1$ means the constant term in $z$, so that for example $[\eta(z)]_1 = \eta_0$. 
Equation \eqref{keytr} contains the key relations in the collective coordinate reformulation of the tRS model.

For completeness, let us mention here that the action of the higher order Hamiltonians $D^{(r)}_{n,\vec{\tau}}$ in terms of bosonic oscillators is given by the operators
\begin{equation}
\mathcal{O}_r(q,t) = \left[ \dfrac{\epsilon_r(z_1, \ldots, z_r)}{\prod_{1 \leqslant i<j \leqslant r} \omega(z_i,z_j)} \eta(z_1) \ldots \eta(z_r) \right]_1\,, \label{Otr}
\end{equation}
where
\begin{equation}
\begin{split}
\omega(z_i,z_j) &= \dfrac{(z_i-q^{-1}z_j)(z_i-tz_j)(z_i-qt^{-1}z_j)}{(z_i-z_j)^3}\,, \\
\epsilon_r(z_1, \ldots, z_r) &= \prod_{1\leqslant i < j \leqslant r} \dfrac{(z_i - t z_j)(z_i-t^{-1}z_j)}{(z_i-z_j)^2}\,.
\end{split}
\end{equation}
We can rewrite the operators using normal ordering as follows
\begin{equation}
\mathcal{O}_r(q,t) = \left[ \prod_{1 \leqslant i<j \leqslant r} \dfrac{(z_i-z_j)^2}{(z_i-qz_j)(z_i-q^{-1}z_j)} :\eta(z_1) \ldots \eta(z_r): \right]_1\,. \label{fco}
\end{equation}
For $r=1$ we immediately recover $\mathcal{O}_1 = [\eta(z)]_1 = \eta_0$. 

\subsubsection{Ding-Iohara Algebra}
Having discussed how to reproduce the action of the tRS Hamiltonians in terms of oscillator modes, let us briefly discuss the relation between the vertex operators we introduced in this section and the free field realization of the algebra known as the quantum Ding-Iohara algebra $\mathcal{U}(q,t)$ \cite{Feigin:2009ab}. Define 
\begin{equation}
g(z) = \dfrac{G^+(z)}{G^-(z)}\,, \;\;\;\;\;\; G^{\pm}(z) = (1-q^{\pm 1}z)(1-t^{\mp 1}z)(1-q^{\mp 1}t^{\pm 1}z)\,.
\end{equation}
Notice that $g(z) = g(z^{-1})^{-1}$. By definition $\mathcal{U}(q,t)$ is the unital associative algebra generated by currents
\begin{equation}
x^{\pm}(z) = \sum_{n \in \mathbb{Z}} x_n^{\pm} z^{-n}\,, \;\;\;\;\;\; \psi^{\pm}(z) = \sum_{\pm n \in \mathbb{N}} \psi_n^{\pm} z^{-n}\,,
\end{equation}
and by central element $\gamma^{\pm 1/2}$ satisfying
\begin{eqnarray}
&& [x^+(z),x^{-}(w)] = \dfrac{(1-q)(1-t^{-1})}{1-qt^{-1}} \left( \delta(\gamma^{-1}z/w) \psi^+(\gamma^{1/2}w) - \delta(\gamma z/w) \psi^-(\gamma^{-1/2}w) \right)\,, \nonumber \\
&& x^{\pm}(z) x^{\pm}(w) = g(z/w)^{\pm 1} x^{\pm}(w) x^{\pm}(z)\,, \nonumber \\
&& \psi^{\pm}(z) \psi^{\pm}(w) = \psi^{\pm}(w) \psi^{\pm}(z)\,, \nonumber \\
&& \psi^{+}(z) \psi^{-}(w) = \dfrac{g (\gamma w/z)}{g(\gamma^{-1}w/z)} \psi^-(w) \psi^+(z)\,, \\
&& \psi^+(z)x^{\pm}(w) = g(\gamma^{\mp 1/2} w/z)^{\mp 1} x^{\pm}(w) \psi^+(z)\,, \nonumber \\
&& \psi^-(z)x^{\pm}(w) = g(\gamma^{\mp 1/2} z/w)^{\pm 1} x^{\pm}(w) \psi^-(z)\,, \nonumber
\end{eqnarray}
where we used the formal expression $\delta(z) = \sum_{m \in \mathbb{Z}} z^m$ for the delta function. The claim, which has been demonstrated in \cite{Feigin:2009ab,2013arXiv1301.4912S,2014SIGMA..10..021S,2013arXiv1309.7094S}, states  that there is a representation $\rho$ of $\mathcal{U}(q,t)$ on the Fock space of Heisenberg algebra \eqref{defHei} given by
\begin{equation}
\rho(\gamma) = \left(tq^{-1}\right)^{1/2}\,, \;\;\;\;\;\; 
\rho(x^+(z)) = \eta(z)\,, \;\;\;\;\;\; 
\rho(x^-(z)) = \xi(z)\,, \;\;\;\;\;\; 
\rho(\psi^{\pm}(z)) = \varphi^{\pm}(z)\,,
\end{equation}
with
\begin{equation}
\begin{split}
\varphi^{+}(z) & = :\eta(\gamma^{1/2}z)\xi(\gamma^{-1/2}z): = \\
& = \text{exp} \left( -\sum_{n>0} \dfrac{1-t^n}{n}(tq^{-1})^{-n/4}(1-(tq^{-1})^{n})a_n z^{-n} \right) = 
\sum_{n \in \mathbb{N}} \varphi^+_n z^{-n}\,, \\
\varphi^{-}(z) &= :\eta(\gamma^{-1/2}z)\xi(\gamma^{1/2}z): \\
& = \text{exp} \left( \sum_{n>0} \dfrac{1-t^{-n}}{n}(tq^{-1})^{-n/4}(1-(tq^{-1})^{n})a_{-n} z^{n} \right) = 
\sum_{n \in \mathbb{N}} \varphi^-_{-n} z^{n}\,.
\end{split}
\end{equation}
An important point to notice is that since $[\varphi^{\pm}(z)]_1 = 1$ we get $[\eta_0,\xi_0] = 0$, which corresponds to the commutativity $[D^{(1)}_{n}(q,t), D^{(1)}_{n}(q^{-1},t^{-1})] = 0$ of the Macdonald operators. 
In the following we shall use Ding-Iohara algebra in the free field realization of the elliptic Ruijsenaars-Schneider model.

\subsection{The Elliptic Ruijsenaars System}\label{EllipticCase}
We can now turn to the collective coordinate description of the eRS model. The goal will be to find an elliptic analogue of the family of commuting operators \eqref{fco} which should represent \eqref{eRSh} (and associated higher order Hamiltonians) in terms of bosonic oscillators.
This problem can be very complicated to solve in a collective field theory approach; nevertheless, as it was done in \cite{Feigin:2009ab}, one can find a way to do this by exploiting the underlying Ding-Iohara algebra structure. Once it is understood that the collective field trigonometric Ruijsenaars-Schneider model is deeply connected to $\mathcal{U}(q,t)$ Ding-Iohara algebra, one can consider its appropriate elliptic deformation $\mathcal{U}(q,t,pq^{-1}t)$.
There are multiple ways to introduce an elliptic deformation to $\mathcal{U}(q,t)$ -- for instance, the proposal by Feigin et.al. \cite{Feigin:2009ab} slightly differs from the work of \cite{2013arXiv1301.4912S,2014SIGMA..10..021S,2013arXiv1309.7094S}. For our purposes the deformation of \cite{Feigin:2009ab} is the most relevant one. In this section we just recollect the main formulas we will need for the upcoming discussion. 

In the elliptic case, the vertex operator gets modified as
\begin{equation}
\eta(z;pq^{-1}t) = \text{exp} \left( \sum_{n>0} \dfrac{1-t^{-n}}{n} \dfrac{1-(pq^{-1}t)^n}{1-p^n} a_{-n}z^n \right) \text{exp} \left( -\sum_{n>0} \dfrac{1-t^n}{n} a_n z^{-n} \right)\,, \label{etap}
\end{equation}
where $p$ is the parameter of elliptic deformation. The elliptic commuting operators $\mathcal{O}_r(q,t;p)$ are constructed from \eqref{etap} as in \eqref{Otr}, with the $\omega$ and $\epsilon_r$ functions replaced by
\begin{equation}
\omega(z_i,z_j;p) = \dfrac{\Theta_p(q^{-1}z_j/z_i)\Theta_p(tz_j/z_i)\Theta_p(qt^{-1}z_j/z_i)}{\Theta_p(z_j/z_i)^3}\,, 
\end{equation}
\begin{equation}
\epsilon_r(z_1, \ldots, z_r;p) = \prod_{1\leqslant i < j \leqslant r} \dfrac{\Theta_p(t z_j/z_i)\Theta_p(t^{-1}z_j/z_i)}{\Theta_p(z_j/z_i)^2}\,.
\end{equation}
The analogue of equation \eqref{keytr}, now relating the eRS Hamiltonian to its bosonized version, reads
\begin{equation*}
\left[ \eta(z;pq^{-1}t) \right]_1 \phi_n(\tau;p) \vert 0 \rangle =  \phi_n(\tau;p)
\left[ t^{-n} \prod_{i=1}^n \dfrac{\Theta_p(qt^{-1}z/\tau_i)}{\Theta_p(qz/\tau_i)} \dfrac{\Theta_p(tz/\tau_i)}{\Theta_p(z/\tau_i)} \eta(z;pq^{-1}t) \right]_1 \vert 0 \rangle 
\end{equation*}
\begin{equation}
+ t^{-n+1}(1-t^{-1})\dfrac{(pt^{-1};p)_{\infty}(ptq^{-1};p)_{\infty}}{(p;p)_{\infty}(pq^{-1};p)_{\infty}} D^{(1)}_{n,\vec{\tau}}(q,t;p) \phi_n(\tau;p) \vert 0 \rangle\,, \label{keyell}
\end{equation}
with $\phi_n(\tau;p) = \phi (\tau_1, \ldots, \tau_n; p)$ the opportune elliptic generalization of \eqref{phi}; see \cite{Feigin:2009ab} for further details. The interesting conjecture of \cite{Feigin:2009ab}, which we will verify in a few cases in the following sections, states that 
\begin{equation}
\lim_{n \rightarrow \infty} \left[ t^{-n} \prod_{i=1}^n \dfrac{\Theta_p(qt^{-1}z/\tau_i)}{\Theta_p(qz/\tau_i)} \dfrac{\Theta_p(tz/\tau_i)}{\Theta_p(z/\tau_i)} \eta(z;pq^{-1}t) \right]_1  \vert 0 \rangle = 0\,. \label{limit}
\end{equation} 
As we shall later see, the limit $n \rightarrow \infty$ allows us to recover information about the finite-difference version of the ILW model starting from the eRS system, and can be intuitively understood as a hydrodynamic limit of eRS. From the gauge theory point of view, this limit will lead to a remarkable relationship between the 5d/3d coupled system of \secref{Ruijsenaarssystemsfromgaugetheory} and the 3d ADHM quiver theory (which, as we will discuss in the next section, is associated to $\Delta$ILW via Bethe/Gauge correspondence).

\section{Gauge Theory Approach to ILW and $\Delta$ILW Systems }\label{ThefinitedifferenceILWsystem}
As we discussed in the previous section trigonometric and elliptic Ruijsenaars-Schneider systems admit a collective field description in terms of $a_m$ modes satisfying a deformed Heisenberg algebra. In \cite{2009JPhA...42N4018S,2009arXiv0911.5005T,2011ntqi.conf..357S} this collective coordinate representation has been interpreted as a realization of the finite-difference version of the Benjamin-Ono and ILW systems respectively ($\Delta$BO and $\Delta$ILW for short); scope of this section is to introduce the main properties of these hydrodynamic systems. The discussion will necessarily be incomplete, since to the best of our knowledge the associated integrable equations have received extremely little attention in the literature; we refer the reader to \cite{2009JPhA...42N4018S,2009arXiv0911.5005T,2011ntqi.conf..357S} for further details. For the sake of clarity, before introducing $\Delta$ILW we will briefly review a few known facts about the standard ILW system and its relation to Calogero models: see also \cite{2005PhRvL..95g6402A,2009JPhA...42m5201A,2014JHEP...07..141B,2015arXiv150507116B}. 

As for the Ruijsenaars models, also the quantum ILW ($\Delta$ILW) system admits a gauge theory description: this time, the associated supersymmetric gauge theory is (via the so-called Bethe/Gauge correspondence \cite{Nekrasov:2009ui,Nekrasov:2009uh}) the 2d (3d) ADHM quiver, as discussed in \cite{2014JHEP...07..141B,2015arXiv150507116B}. At the end of this section we will recollect the main points of this correspondence, focussing on how it is possible to recover the quantum ILW ($\Delta$ILW) spectrum by studying the Coulomb branch vacua of the ADHM quiver theory; further details on the ADHM theory are presented in Appendix \ref{appA}.

\subsection{The ILW system}
\label{Sec:ILWDescription}
\begin{figure}[h]
\centering
\includegraphics[width=0.8\textwidth]{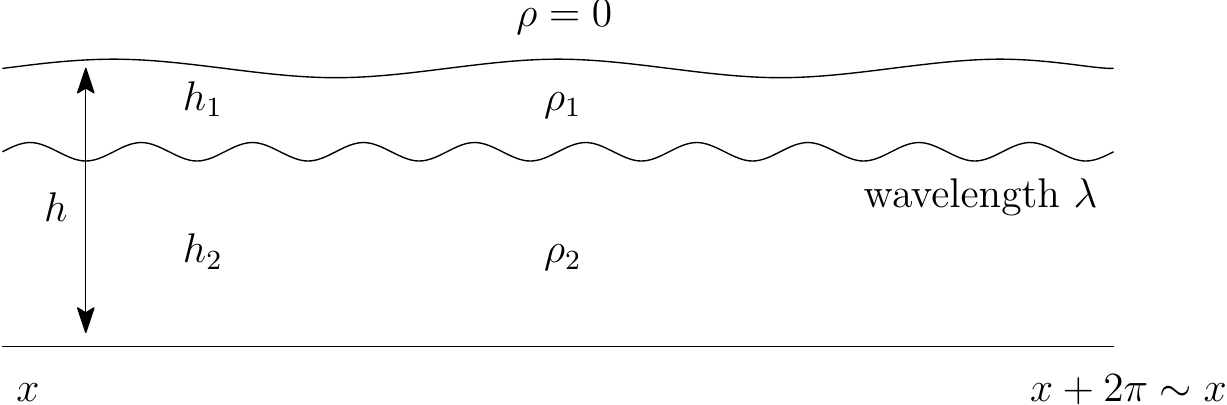} 
\caption{The classical periodic Intermediate Long Wave system} \label{fig:figILW}
\end{figure}

Consider the situation illustrated in Figure \ref{fig:figILW}. We have a system of two fluids of densities $\rho_1 < \rho_2$ and depths $h_1 < h_2$ in a periodic (period $L = 2 \pi$) channel of total depth $h = h_1 + h_2$. \\

The classical periodic Intermediate Long Wave system describes the propagation of waves in the interface of the two fluids, due to gravitational effects \cite{opac-b1122380}. There are three possible regimes, according to the value of the ratio $\delta = \frac{h}{\lambda}$ with $\lambda$ typical wavelength:
\begin{itemize}
\item $h \ll \lambda$, long wave: Korteweg-de Vries (KdV) regime for $\delta \rightarrow 0$
\item $h \gg \lambda$, short wave: Benjamin-Ono (BO) regime for $\delta \rightarrow \infty$
\item $h \sim \lambda$, intermediate wave: Intermediate Long Wave (ILW) regime for $\delta \sim 1$
\end{itemize}
The evolution of the wave profile $u(x,t)$ is determined by the classical ILW partial integro-differential equation
\begin{equation}
u_t = 2 u_{xx}- i \beta \partial^2_x u^{H} \label{ILWeq}
\end{equation}
with $\beta$ a parameter depending on the densities $\rho_{1,2}$, the depths $h_{1,2}$ and the standard gravity constant $g$. 
Here $u^{H}$ is the Hilbert transformed wave-function
\begin{equation}
u^H = \dfrac{1}{2 \pi} P.V. \int_0^{2\pi} \zeta(y-x;\widetilde{p}) u(y) dy\,, \label{Wei}
\end{equation} 
where $\zeta$ is the Weierstrass zeta function and we defined $\widetilde{p} = e^{-2 \pi \delta}$. The zeta function reduces to the cotangent function in the limit $\widetilde{p} \rightarrow 0$, giving the Benjamin-Ono equation
\begin{equation}
u_t = 2 u_{xx}- i \beta \partial^2_x u^{H}\,, \quad u^H = \dfrac{1}{2 \pi} P.V. \int_0^{2\pi} \cot(y-x) u(y) dy \,,
\end{equation} 
while in the $\widetilde{p} \rightarrow 1$ limit we recover from \eqref{ILWeq} the KdV equation
\begin{equation}
u_t = 2 u u_x + \dfrac{\beta}{3} u_{xxx}\,.
\end{equation}
The key characteristic of the ILW equation \eqref{ILWeq} is that it is \textit{integrable}: if we introduce the Poisson bracket
\begin{equation}
\{ u(x), u(y) \} = \delta'(x-y) \label{PBstr}
\end{equation}
one can show that there exist an infinite number of linearly independent conserved quantities (Hamiltonians) $I_l$ which are in involution with respect to this Poisson bracket
\begin{equation}
\{ I_l, I_m \} = 0\,.
\end{equation} 
The first conserved quantities read 
\begin{equation}
I_1 = \int \left[\frac{1}{2} u^2 \right] dx\,, \;\;\;\; I_2 = \int \left[ \frac{1}{3} u^3 + i \frac{\beta}{2} u u^{H}_x \right] dx\,, \;\;\;\; \ldots
\end{equation}
The ILW equation is determined by the Hamiltonian $I_2$ via
\begin{equation}
u_t = \{ u, I_2 \}\,.
\end{equation}
In a sense, integrability for a partial differential equation can be interpreted as an extension of the usual definition of classical Liouville integrability in the case of a system with an infinite number of degrees of freedom. What is peculiar with respect to the Liouville case is that integrability of a partial differential equation implies the existence of an infinite number of exact solutions, known as $n$-\textit{soliton} solutions: very roughly, they are waves whose profile does not change with time, apart from the instants in which two solitons collide. 

\subsubsection{Solitonic Solutions as Particle Systems}
The class of $n$-soliton solutions provide a way to relate integrable partial differential equations to integrable systems with a finite number $n$ of degrees of freedom; for the ILW case, the associated system is the $n$-particle Calogero-Sutherland model. To clarify this point, let us consider the simple example of the non-periodic ($L \rightarrow \infty$) BO system ($\delta \rightarrow \infty$) given by the equation
\begin{equation}
u_t = 2 u_{xx}- i \beta \partial^2_x u^{H}\,, \;\;\;\;\;\; u^H = P.V. \int_0^{\infty} \dfrac{u(y)}{y-x} dy\,. \label{npBO}
\end{equation} 
A $n$-soliton solution can be written in terms of the \textit{pole ansatz}
\begin{equation}
u(x,t) = \sum_{j=1}^n \left( \dfrac{i \beta}{x - a_j(t)} - \dfrac{i \beta}{x - a^*_j(t)} \right)\,, \label{poleansatz}
\end{equation}
where the poles $a_j(t)$ determine the positions of the solitons and $a^*_j(t)$ is its complex conjugate. Inserting \eqref{poleansatz} into \eqref{npBO} one finds that this Ansatz is a solution to the non-periodic BO equation if and only if the poles dynamics satisfies the equations of motion 
\begin{equation}
\ddot{a}_j = \sum_{l \neq j}^n \dfrac{2 \beta^2}{(a_j - a_l)^3}\,.
\end{equation}
These are the same equations of motion arising from an $n$-particle classical \textit{rational Calogero} system, which is a classical integrable system of $n$ particles on a line interacting via the Hamiltonian
\begin{equation}
\mathcal{H}^{(n)}_{rCS} = \dfrac{1}{2} \sum_{j=1}^n p_j^2 + \sum_{l<j}^n \dfrac{\beta^2}{(a_l - a_j)^2}\,. \label{rCS}
\end{equation}
A similar analysis is valid for the periodic BO and ILW systems, whose $n$-solitons are associated respectively to the $n$-particle \textit{trigonometric} and \textit{elliptic Calogero-Sutherland} models; clearly, the pole ansatz in these cases will be an obvious trigonometric or elliptic generalization of \eqref{poleansatz}. 

\subsubsection{Quantization}
At the quantum level, the \textit{quantum} ILW or BO equations can be obtained from the solitons solutions simply by considering the quantum versions of Calogero systems: for example, the quantum version of \eqref{rCS} reads
\begin{equation}
\mathcal{H}^{(n)}_{rCS} = -\dfrac{1}{2} \sum_{j=1}^n \dfrac{d^2}{da_j^2} + \sum_{l<j}^n \dfrac{\beta(\beta - 1)}{(a_l - a_j)^2}\,, \label{rCS}
\end{equation}
and gives rise to the equations of motion 
\begin{equation}
\ddot{a}_j = \sum_{l \neq j}^n \dfrac{2 \beta(\beta - 1)}{(a_j - a_l)^3}
\end{equation}
for the vacuum expectation values of the $a_j$. The pole ansatz \eqref{poleansatz} will satisfy the equation
\begin{equation}
u_t = 2 u_{xx}- i (\beta + \beta^{-1}) \partial^2_x u^{H}\,, \;\;\;\;\;\; u^H = P.V. \int_0^{\infty} \dfrac{u(y)}{y-x} dy \,,
\end{equation} 
which is therefore called the non-periodic quantum BO equation. In the same way, the periodic quantum ILW equation will be given by
\begin{equation}
u_t = 2 u_{xx}- i (\beta + \beta^{-1}) \partial^2_x u^{H} \label{ILWeqqu}
\end{equation}
with $u^H$ as in \eqref{Wei}, and similarly for the periodic quantum BO equation. The main difference between \eqref{ILWeq} and \eqref{ILWeqqu} is the replacement $\beta \rightarrow \beta + \beta^{-1}$. \\

In what follows the quantum ILW system will be obtained from \eqref{ILWeqqu} after quantization of Poisson structure \eqref{PBstr}. The idea is the following. One starts by expanding the $u(x)$ field in Fourier modes:
\begin{equation}
u(x) = \sum_{\substack{m \in \mathbb{Z} \\ m \neq 0}} a_m e^{i m x}\,.
\end{equation}
The Poisson bracket \eqref{PBstr} implies that the Fourier modes $a_m$ satisfy
\begin{equation}
\{ a_m, a_{-n} \} = - i m \delta_{m,n}\,.
\end{equation}
We now promote the Poisson bracket to a quantum commutator and the Fourier modes to quantum operators, thus obtaining the Heisenberg algebra
\begin{equation}
[ a_m, a_{-n} ] = \hbar m \delta_{m,n}\,. \label{HA}
\end{equation}
Planck constant $\hbar$  will be often put to unity in the following. One needs now to understand what are the quantum Hamiltonians $\widehat{I}_l$ that characterize the quantum ILW system. In order to see this we can simply take the normal ordered product of the operators, i.e. $\widehat{I}_l \,=\, :I_l:$. These operators do not commute under \eqref{HA}. We therefore need to add $o(\hbar)$ corrections in order to ensure commutativity:
\begin{equation}
\widehat{I}_l \,=\, :I_l: + o(\hbar) \;\;\; \text{such that} \;\;\; [\widehat{I}_l, \widehat{I}_m] = 0\,.
\end{equation}
Unfortunately, only the first few quantum Hamiltonians are known for the ILW system; for example we have
\begin{equation} 
\widehat{I}_2 = \sum_{m>0} a_{-m}a_m \label{ILW2}\,,
\end{equation}
\begin{equation}
\widehat{I}_3 = i \dfrac{\beta + \beta^{-1}}{2} \sum_{m>0} m \dfrac{1 + (-\tilde{p})^m}{1 - (-\tilde{p})^m} \, a_{-m} a_m + \dfrac{1}{2} \sum_{m,n > 0} (a_{-m-n} a_m a_n + a_{-m} a_{-n} a_{m+n})\,. \label{ILW3}
\end{equation}
The quantum ILW problem can therefore be stated as finding the whole set of commuting quantum ILW Hamiltonians, as well as their eigenstates and eigenvalues. The solution to this problem is only known in the BO limit; for the generic ILW system far less is known. Nevertheless steps towards this direction were made in \cite{2014JHEP...07..141B}, where it was shown that the ILW spectrum can be computed (perturbatively in $\tilde{p} \sim 0$ around the known BO solution) by studying the Coulomb branch of the 2d ADHM quiver theory. The above correspondence with the gauge theory will be shortly reviewed in the end of this section.

\subsection{The $\Delta$ILW System and its Spectrum} 
\label{Sec:somesec}  

In \cite{2009JPhA...42N4018S} the authors introduced and discussed in some detail a finite-difference version of the classical ILW equation. This reads as follows
\begin{equation}
\dfrac{\partial}{\partial t_0} \eta (z,t_0) = \dfrac{i}{2} \eta (z,t_0) P.V. \int_{-1/2}^{1/2} 
(\Delta_{\gamma} \zeta)(\pi (w-z)) \cdot \eta(w,t_0) dw\,. \label{fde}
\end{equation}
Here the discrete Laplacian $\Delta_{\gamma}$ is defined as $(\Delta_{\gamma}f)(x) = f(x + \gamma) - 2f(x) + f(x - \gamma)$ and $\gamma$ is a complex number. It is easy to show that in the limit $\gamma \rightarrow 0$ \eqref{fde} reduces to \eqref{ILWeq}, after an appropriate Galilean transformation on $\eta(z, t_0)$. 
The finite-difference Benjamin-Ono limit of this equation has been studied in greater detail in \cite{2009arXiv0911.5005T,2011ntqi.conf..357S}, both at the classical and the quantum level. 

Based on our results we expect that the quantum $\Delta$ILW system to have a deep connection to the quantum elliptic Ruijsenaars-Schneider model and the elliptic deformation of the Ding-Iohara algebra which we discussed in \secref{FreefieldrealizationofRuijsenaars}. Since classical $\Delta$ILW Hamiltonians $\mathcal{H}_r$ given in \cite{2009JPhA...42N4018S} can be exactly reproduced in a certain limit of commuting operators $\mathcal{O}_r$ introduced in \secref{EllipticCase}; we propose that our operators $\mathcal{O}_r$ coincide with quantum $\Delta$ILW Hamiltonians $\widehat{\mathcal{H}}_r$. Moreover $\eta(z;pq^{-1}t)$ field of \eqref{etap} can be shown to satisfy \eqref{fde} in the classical limit, where the Hamiltonian generating the time evolution of the system is $\mathcal{H}_1$. 

In order to see the correspondence between the $\Delta$ILW system and the elliptic Ruijsenaars-Schneider model, we will start by computing here the spectrum of the $\Delta$ILW system. 
First, as we will see in Section \ref{5.7.2}, elliptic deformation parameter $p$ of the Ruijsenaars system and ILW parameter $\tilde{p} = e^{-2\pi \delta}$ need to be identified as\footnote{In terms of quantum cohomology, parameter $\delta$ coincides with the K\"{a}hler modulus of $\mathcal{M}_{k,1}$ and is the same as parameter $t$ which was used in \cite{2014JHEP...07..141B}.}
\begin{equation}
p = -\tilde{p} \sqrt{qt^{-1}}\,. \label{anticipation}
\end{equation}
Moreover, we shall substitute $q$ and $t$ by $e^{i \gamma \epsilon_1}$ and $e^{- i \gamma \epsilon_2}$ respectively in order to make contact with the gauge theory results of the following Section \ref{section 5.6}, and rewrite \eqref{etap} as
\begin{equation}
\eta(z;pq^{-1}t) = \text{exp}\left(\sum_{n>0}\lambda_{-n}z^n\right) \text{exp}\left(\sum_{n>0}\lambda_{n}z^{-n}\right)\,, 
\end{equation}
with commutation relations for the $\lambda_m$
\begin{equation}
[\lambda_m, \lambda_n] = -\dfrac{1}{m} \dfrac{(1-q^m)(1-t^{-m})(1-(pq^{-1}t)^m)}{1-p^m} \delta_{m+n,0}\,.
\end{equation} 
We now have to consider the eigenvalue problem for the first $\Delta$ILW quantum Hamiltonian 
\begin{equation}
\widehat{\mathcal{H}}_1 = \left[\eta\left(z; -\tilde{p}q^{-\frac{1}{2}} t^{\frac{1}{2}}\right) \right]_1 .
\end{equation} 
Denoting by $k$ the number of solitons present in the $\Delta$ILW solution, we restrict ourselves to the cases with $k$ up to $3$ in the following. 

\subsubsection{Absence of solitons} 

The state corresponding to $k=0$ is just the vacuum state $\vert 0 \rangle$, for which
\begin{equation}
\left[\eta\left(z; -\tilde{p}q^{-\frac{1}{2}} t^{\frac{1}{2}}\right) \right]_1 \vert 0 \rangle = \vert 0 \rangle = \mathcal{E}_1 \vert 0 \rangle
\end{equation}
We can therefore conclude that
\begin{equation}
\mathcal{E}_1 = 1 \label{(0osc)}\,.
\end{equation}

\subsubsection{One soliton} 

The generic $k=1$ state can be expressed as
\begin{equation}
c_1 \lambda_{-1} \vert 0 \rangle
\end{equation}
and depends on a normalization constant $c_1$ which is not relevant for our purpose of computing its eigenvalue. The eigenvalue equation in this case reduces to 
\begin{equation}
\left[\eta\left(z; -\tilde{p}q^{-\frac{1}{2}} t^{\frac{1}{2}}\right) \right]_1 c_1 \lambda_{-1} \vert 0 \rangle 
= \big[ 1 + \lambda_{-1} \lambda_1 \big] c_1 \lambda_{-1} \vert 0 \rangle
= \mathcal{E}_1 c_1 \lambda_{-1} \vert 0 \rangle
\end{equation}
from which we obtain
\begin{equation}
\begin{split}
\mathcal{E}_1 \,=\, & (q + t^{-1} - qt^{-1}) - \dfrac{(1-q)(1-t)(q-t)}{qt} 
\dfrac{\tilde{p}\sqrt{q t^{-1}}}{1+\tilde{p}\sqrt{q t^{-1}}} \label{(1osc)}
\end{split}
\end{equation}
exact in $\tilde{p}$.

\subsubsection{Two solitons}
 
A state with $k=2$ can generically be written as
\begin{equation}
(c_1 \lambda_{-1}^2 + c_2 \lambda_{-2}) \vert 0 \rangle
\end{equation}
in terms of two constants $c_1$, $c_2$ which are to be determined. The eigenvalue equation 
\begin{equation}
\begin{split}
& \left[\eta\left(z; -\tilde{p}q^{-\frac{1}{2}} t^{\frac{1}{2}}\right) \right]_1 (c_1 \lambda_{-1}^2 + c_2 \lambda_{-2}) \vert 0 \rangle  
= \mathcal{E}_1 (c_1 \lambda_{-1}^2 + c_2 \lambda_{-2}) \vert 0 \rangle = \\
& = \left[ 1 + \lambda_{-1} \lambda_1 + \lambda_{-2} \lambda_2 + \dfrac{1}{2}\left(\lambda_{-2}\lambda_1^2 + \lambda_{-1}^2 \lambda_2 \right) + \dfrac{1}{4} \lambda_{-1}^2 \lambda_1^2 \right] (c_1 \lambda_{-1}^2 + c_2 \lambda_{-2}) \vert 0 \rangle
\end{split}
\end{equation}
has the two solutions
\begin{equation}
\begin{split}
\mathcal{E}_1^{(1)} \,=\, & (q^2 + t^{-1} - q^2t^{-1}) - \tilde{p} \sqrt{q t^{-1}} \dfrac{(1-q^2)(1-t)^2(q-t)}{t(1-qt)} \\
& + \tilde{p}^2 \dfrac{(1-q^2)(1-t)(q-t)}{q t^2(1-qt)^3} [ q^3+t+qt+q^2t^2+3q^3t^2+q^4t^2+2q^2t^3 \\
& \hspace{4.5 cm} - 3q^2t-2q^3t-2qt^2-qt^3-2q^4t^3 ] + o(\tilde{p}^3) \label{20osc} 
\end{split} 
\end{equation}
\begin{equation}
\begin{split}
\hspace*{-0.5 cm} \mathcal{E}_1^{(2)} \,=\, & (q + t^{-2} - qt^{-2}) - \tilde{p} \sqrt{q t^{-1}} \dfrac{(1-q)^2(1-t^2)(q-t)}{qt^2(1-qt)} \\
& + \tilde{p}^2 \dfrac{(1-q)(1-t^2)(q-t)}{t^3(1-qt)^3} [ 2+2q^2t+3q^2t^2+t^3+2qt^3-q-3qt \\ 
& \hspace{4.5 cm} -q^3t-2t^2-qt^2-q^2t^3-q^2t^4 ] + o(\tilde{p}^3) \label{11osc}
\end{split} 
\end{equation}
related by the exchange $q \longleftrightarrow t^{-1}$, i.e. $\e_1 \longleftrightarrow \e_2$.
We therefore have two eigenstates, whose constants $c_1$, $c_2$ have to satisfy the relations
\begin{equation}
\begin{split}
& c_2 = \left( -2 \dfrac{1-q}{1+q} + \dfrac{\tilde{p}}{\sqrt{qt}} \dfrac{4(1-q)(q-t)}{(1+q)(1-qt)} + o(\tilde{p}^2) \right) c_1 \\
& c_2 = \left( -2 \dfrac{1-t^{-1}}{1+t^{-1}} + \tilde{p}\sqrt{qt} \, \dfrac{4(1-t)(q-t)}{(1+t)(1-qt)} + o(\tilde{p}^2) \right) c_1
\end{split} 
\end{equation}
The remaining constant $c_1$ only enters in the normalization of the eigenstates, and will be of no importance for our discussion. 

\subsubsection{Three solitons}

A generic state with $k=3$ can be written as
\begin{equation}
(c_1 \lambda_{-1}^3 + c_2 \lambda_{-2}\lambda_{-1} + c_3 \lambda_{-3})\vert 0 \rangle
\end{equation}
The eigenvalue equation 
\begin{equation}
\begin{split}
& \left[\eta\left(z; -\tilde{p}q^{-\frac{1}{2}} t^{\frac{1}{2}}\right) \right]_1 (c_1 \lambda_{-1}^3 + c_2 \lambda_{-2}\lambda_{-1} + c_3 \lambda_{-3})\vert 0 \rangle \\
& = \mathcal{E}_1 (c_1 \lambda_{-1}^3 + c_2 \lambda_{-2}\lambda_{-1} + c_3 \lambda_{-3})\vert 0 \rangle \\
& = \Big[ 1 + \lambda_{-1} \lambda_1 + \lambda_{-2} \lambda_2 + \lambda_{-3}\lambda_3  
+ \dfrac{1}{2}\left(\lambda_{-2}\lambda_1^2 + \lambda_{-1}^2 \lambda_2 + 2\lambda_{-3}\lambda_2 \lambda_1 + 2 \lambda_{-1}\lambda_{-2}\lambda_3 \right)  \\
&\;\;\;\;\; + \dfrac{1}{4} \lambda_{-1}^2 \lambda_1^2 + \lambda_{-1}\lambda_{-2}\lambda_1 \lambda_2 + \dfrac{1}{6}\left( \lambda_{-1}^3 \lambda_3 + \lambda_{-3}\lambda_1^3 + \lambda_{-1}^3 \lambda_1 \lambda_2 + \lambda_{-1} \lambda_{-2} \lambda_1^3 \right) \\
&\;\;\;\;\; + \dfrac{1}{36} \lambda_{-1}^3 \lambda_1^3 \Big] 
(c_1 \lambda_{-1}^3 + c_2 \lambda_{-2}\lambda_{-1} + c_3 \lambda_{-3})\vert 0 \rangle
\end{split}
\end{equation}
leads to an equation for eigenvalue $E_3$ with three solutions
\begin{equation}
\begin{split}
\mathcal{E}_1^{(1)} \,=\, & (q^3 + t^{-1} - q^3t^{-1}) - \tilde{p} \sqrt{q t^{-1}} \dfrac{q(1-t)^2(1-q^3)(q-t)}{t(1-q^2t)} \\
& + \tilde{p}^2 \dfrac{(1-t)^2(1-q^3)(q-t)}{t^2(1-q^2t)^3} [ q^4+t+2qt+q^5t+qt^2+q^5t^2+2q^6t^2 \\
& \hspace{4.5 cm} - q^2t-3q^3t-2q^4t-2q^3t^2-q^4t^2 ] + o(\tilde{p}^3) \label{300osc}
\end{split}
\end{equation}
\begin{equation}
\begin{split}
\mathcal{E}_1^{(2)} \,=\, & (q^2 + qt^{-1} + t^{-2} - qt^{-2} -q^2t^{-1}) \\
& - \tilde{p} \sqrt{q t^{-1}} \dfrac{(1-q)(1-t)(q-t)}{qt^2(1-qt^2)(1-q^2t)} [ 1+2qt+2q^2t^2+2q^3t^3+q^4t^4 \\
& \hspace{4.5 cm} -q^2-q^3t-2qt^2-q^4t^2-qt^3-2q^2t^3 ] + o(\tilde{p}^2) \label{210osc}
\end{split}
\end{equation}
\begin{equation}
\begin{split}
\mathcal{E}_1^{(3)} \,=\, & (q + t^{-3} - qt^{-3}) - \tilde{p} \sqrt{q t^{-1}} \dfrac{(1-q)^2(1-t^3)(q-t)}{qt^3(1-qt^2)} \\
& + \tilde{p}^2 \dfrac{(1-q)^2(1-t^3)(q-t)}{t^4(1-qt^2)^3} [ 2+t+qt+q^2t^2+t^5+2qt^5+qt^6 \\
& \hspace{4.5 cm} - t^2-2qt^2-2t^3-3qt^3-qt^4 ] + o(\tilde{p}^3) \label{111osc}
\end{split}
\end{equation}
We conclude that there are three eigenstates; again, it is possible to determine $c_2$ and $c_3$ in terms of the overall normalization $c_1$ starting from the eigenvalue equations, as we showed in the previous case.

\subsection{The $\Delta$ILW Spectrum from Gauge Theory} \label{section 5.6}

Although the procedure described above provides the $\widehat{H}_1$ eigenvalue at specified $k$, it turns out that it is possible to obtain the same results from gauge theory, more precisely from the so-called ADHM quiver gauge theory in two or three dimensions. 
  
The relation between the ADHM gauged linear sigma model for the $U(1)$ theory ($N=1$ model) and the quantum ILW system has been discussed in terms of Bethe/Gauge correspondence in \cite{2014JHEP...07..141B}. There the authors explained why the equations which determine supersymmetric vacua in the Coulomb branch of the 2d ADHM theory correspond to the Bethe Ansatz Equations for ILW, as well as how the local gauge theory observables $\langle \text{Tr}\,\Sigma^l \rangle$ evaluated at the solutions of these equations give the ILW spectrum.  
Here we propose a similar correspondence to hold between the $N=1$ ADHM theory on $\Complex \times S^1_{\gamma}$ and quantum $\Delta$ILW. We shall provide the calculations supporting this statement below, while later in \secref{Sec:StringThDer} we shall explain how the ADHM theory arises in our construction by using string theory dualities. 

When the radius of the circle $\gamma$ is small the infrared description of the sigma model is effectively two-dimensional. The supersymmetric Coulomb branch vacua equations for $N=1$ will be (see Appendix \ref{appA})
\begin{equation}
\begin{split}
& 
\sin [\frac{\gamma}{2} (\Sigma_s - a)]
\prod_{\substack{t = 1 \\ t\neq s}}^k \dfrac{\sin [\frac{\gamma}{2} (\Sigma_{st} - \epsilon_1)] \sin [\frac{\gamma}{2} (\Sigma_{st} - \epsilon_2)]}{\sin [\frac{\gamma}{2} ( \Sigma_{st})] \sin [\frac{\gamma}{2} (\Sigma_{st} - \epsilon)]} = \\
& \tilde{p} \, \sin [\frac{\gamma}{2} ( -\Sigma_s + a - \epsilon)] 
\prod_{\substack{t = 1 \\ t\neq s}}^k \dfrac{\sin [\frac{\gamma}{2} ( \Sigma_{st} + \epsilon_1)] \sin [\frac{\gamma}{2} (\Sigma_{st} + \epsilon_2)]}{\sin [\frac{\gamma}{2} ( \Sigma_{st})] \sin [\frac{\gamma}{2} (\Sigma_{st} + \epsilon)]} \label{BAE}
\end{split}
\end{equation}
because of the 1-loop contributions coming from the KK tower of chiral multiplets\footnote{Equations \eqref{BAE} reduce to the Bethe Ansatz Equations for quantum ILW of \cite{2014JHEP...07..141B} when $\gamma \rightarrow 0$.}. Here $\epsilon = \epsilon_1 + \epsilon_2$ and $\tilde{p} = e^{-2 \pi \xi}$ with $\xi$ Fayet-Iliopoulos parameter of the ADHM theory\footnote{As discussed in \cite{2014JHEP...07..141B}, the Fayet-Iliopoulos parameter $\xi$ coincides with the ILW parameter $\delta$ previously introduced.}. For simplicity, from now on we will set $a=0$. When $\xi \rightarrow \infty$ (i.e. $\tilde{p} \rightarrow 0$), the solutions are labelled by partitions $\lambda$ of $k$, and are given by 
\begin{equation}
\Sigma_s = (i-1)\e_1 + (j-1)\e_2 \;\;\;\text{mod } 2 \pi i \label{solBO}
\end{equation} 
\begin{figure}[h]
\centering
\includegraphics[width=0.15\textwidth]{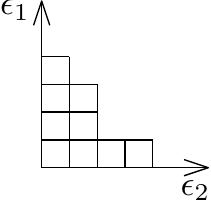}
\caption{The partition (4,3,1,1) of $k=9$} 
\end{figure}

For finite $\xi$ we can change variables to $\sigma_s = e^{i \gamma \Sigma_s}$, $q = e^{i \gamma \epsilon_1}$, $t = e^{-i \gamma \epsilon_2}$ and rewrite \eqref{BAE} as 
\begin{equation}
\begin{split}
& (\sigma_s - 1) \prod_{\substack{t = 1 \\ t\neq s}}^k \dfrac{(\sigma_{s} - q \sigma_t) (\sigma_{s} - t^{-1} \sigma_t)}{( \sigma_{s} - \sigma_t) (\sigma_{s} - q t^{-1} \sigma_{t})} =
-\tilde{p}\sqrt{q t^{-1}} \, (\sigma_s - q^{-1} t) \prod_{\substack{t = 1 \\ t\neq s}}^k \dfrac{(\sigma_{s} - q^{-1} \sigma_t) (\sigma_{s} - t \sigma_t)}{(\sigma_{s} - \sigma_t) (\sigma_{s} - q^{-1} t \sigma_{t})}\,. \label{BAE2}
\end{split}
\end{equation}
Perturbatively in small $\tilde{p}$, solutions to \eqref{BAE2} are still labelled by partitions $\lambda$ of $k$.
We propose \eqref{BAE2} to be the Bethe Ansatz Equations for quantum $\Delta$ILW: therefore the $\Delta$ILW eigenfunctions will be in one-to-one correspondence with partitions $\lambda$, and the eigenvalues of the quantum $\Delta$ILW Hamiltonians $\widehat{\mathcal{H}}_r$ will be related to the local 3d gauge theory observables $\langle \text{Tr} \, \sigma^r \rangle$ evaluated at solutions $\lambda$ of \eqref{BAE2}. In particular, from what we noticed earlier, we expect the equivariant Chern character of the universal $U(1)$ bundle over the instanton moduli space
\begin{equation}
\mathcal{E}_1^{(\lambda)} = 1 - (1-q)(1-t^{-1}) \sum_s \sigma_s \Big\vert_{\lambda} \label{chern}
\end{equation} 
to be the eigenvalue of $\widehat{\mathcal{H}}_1$. Let us remark here that it is easy to show that $\mathcal{E}_1$ is a generating function for the eigenvalues $E_l$ of the quantum ILW Hamiltonians $\widehat{I}_l$ according to
\begin{equation}
\mathcal{E}_1^{(\lambda)} = 1 + \gamma^2 \e_1 \e_2 k + \gamma^3 \e_1 \e_2 E^{(\lambda)}_3 + \gamma^4 \e_1 \e_2 E^{(\lambda)}_4 + \ldots \label{expansion}
\end{equation}
In the following we will list the eigenvalue $\mathcal{E}_1^{(\lambda)}$ for the solutions of \eqref{BAE2} at low $k$.

\subsubsection{Absence of solitons}

In this case $k=0$ and the solution is trivial
\begin{equation}
\mathcal{E}_1^{(\varnothing)} = 1 \label{(0)}
\end{equation}
and coincides with \eqref{(0osc)}.

\subsubsection{One soliton}
For $k=1$ the solution can be computed exactly in $\tilde{p}$ and is given by
\begin{equation}
\begin{split}
\mathcal{E}_1 \,=\, & (q + t^{-1} - qt^{-1}) - \dfrac{(1-q)(1-t)(q-t)}{qt} 
\dfrac{\tilde{p}\sqrt{q t^{-1}}}{1+\tilde{p}\sqrt{q t^{-1}}} \label{(1)}\,.
\end{split}
\end{equation}
This coincides with \eqref{(1osc)} thanks to the identification $p = - \tilde{p} \sqrt{qt^{-1}}$.

\subsubsection{Two solitons}
For $k=2$ there are two solutions, labelled by the two partitions of $k$.
\begin{itemize}
\item $k=2$, partition $(2,0)$ \\
\begin{equation}
\begin{split}
\mathcal{E}_1^{(2,0)} \,=\, & (q^2 + t^{-1} - q^2t^{-1}) - \tilde{p} \sqrt{q t^{-1}} \dfrac{(1-q^2)(1-t)^2(q-t)}{t(1-qt)} \\
& + \tilde{p}^2 \dfrac{(1-q^2)(1-t)(q-t)}{q t^2(1-qt)^3} [ q^3+t+qt+q^2t^2+3q^3t^2+q^4t^2+2q^2t^3 \\
& \hspace{4.5 cm} - 3q^2t-2q^3t-2qt^2-qt^3-2q^4t^3 ] + o(\tilde{p}^3) \label{20}
\end{split}
\end{equation}
This expression reproduces \eqref{20osc}.
\item $k=2$, partition $(1,1)$ \\
\begin{equation}
\begin{split}
\mathcal{E}_1^{(1,1)} \,=\, & (q + t^{-2} - qt^{-2}) - \tilde{p} \sqrt{q t^{-1}} \dfrac{(1-q)^2(1-t^2)(q-t)}{qt^2(1-qt)} \\
& + \tilde{p}^2 \dfrac{(1-q)(1-t^2)(q-t)}{t^3(1-qt)^3} [ 2+2q^2t+3q^2t^2+t^3+2qt^3-q-3qt \\ 
& \hspace{4.5 cm} -q^3t-2t^2-qt^2-q^2t^3-q^2t^4 ] + o(\tilde{p}^3) \label{11}
\end{split}
\end{equation}
This coincides with \eqref{11osc}.

\subsubsection{Three solitons}

For $k = 3$ there are three solutions, labelled by the three partitions of $k$.
\item $k=3$, partition $(3,0,0)$ \\
\begin{equation}
\begin{split}
\mathcal{E}_1^{(3,0,0)} \,=\, & (q^3 + t^{-1} - q^3t^{-1}) - \tilde{p} \sqrt{q t^{-1}} \dfrac{q(1-t)^2(1-q^3)(q-t)}{t(1-q^2t)} \\
& + \tilde{p}^2 \dfrac{(1-t)^2(1-q^3)(q-t)}{t^2(1-q^2t)^3} [ q^4+t+2qt+q^5t+qt^2+q^5t^2+2q^6t^2 \\
& \hspace{4.5 cm} - q^2t-3q^3t-2q^4t-2q^3t^2-q^4t^2 ] + o(\tilde{p}^3) \label{300}
\end{split}
\end{equation}
This reproduces the result of \eqref{300osc}.
\item $k=3$, partition $(2,1,0)$ \\
\begin{equation}
\begin{split}
\mathcal{E}_1^{(2,1,0)} \,=\, & (q^2 + qt^{-1} + t^{-2} - qt^{-2} -q^2t^{-1}) \\
& - \tilde{p} \sqrt{q t^{-1}} \dfrac{(1-q)(1-t)(q-t)}{qt^2(1-qt^2)(1-q^2t)} [ 1+2qt+2q^2t^2+2q^3t^3+q^4t^4 \\
& \hspace{4.5 cm} -q^2-q^3t-2qt^2-q^4t^2-qt^3-2q^2t^3 ] + o(\tilde{p}^2) \label{210}
\end{split}
\end{equation}
This reproduces the result of \eqref{210osc}.
\item $k=3$, partition $(1,1,1)$ \\
\begin{equation}
\begin{split}
\mathcal{E}_1^{(1,1,1)} \,=\, & (q + t^{-3} - qt^{-3}) - \tilde{p} \sqrt{q t^{-1}} \dfrac{(1-q)^2(1-t^3)(q-t)}{qt^3(1-qt^2)} \\
& + \tilde{p}^2 \dfrac{(1-q)^2(1-t^3)(q-t)}{t^4(1-qt^2)^3} [ 2+t+qt+q^2t^2+t^5+2qt^5+qt^6 \\
& \hspace{4.5 cm} - t^2-2qt^2-2t^3-3qt^3-qt^4 ] + o(\tilde{p}^3) \label{111}
\end{split}
\end{equation}
This reproduces the result of \eqref{111osc}.
\end{itemize}
These computations justify our proposal for the ADHM theory on $\Complex \times S^1_{\gamma}$ to be the gauge theory whose underlying integrable system corresponds to $\Delta$ILW, in a natural generalization of the two-dimensional setup.

\section{$\Delta$ILW as Collective Coordinate Elliptic Ruijsenaars-Schneider Model} \label{section5} 
Let us summarize what we have done so far. In \secref{Ruijsenaarssystemsfromgaugetheory} we introduced the $n$-particles quantum trigonometric and elliptic Ruijsenaars-Schneider models using the Seiberg-Witten description of the 5d $\mathcal{N}=1^*$ $U(n)$ gauge theory with defects. The we constructed the eigenfunctions and the eigenvalues for the tRS or eRS models. We were able to perform explicit computations for the eRS system, thanks to our understanding of instanton configurations in supersymmetric gauge theories. In \secref{FreefieldrealizationofRuijsenaars} we reviewed the collective coordinate realization of tRS and eRS in terms of free bosons; in \secref{ThefinitedifferenceILWsystem} this realization has been given an interpretation in terms of a finite-difference version of the Benjamin-Ono and the ILW systems, which from the gauge theory point of view are related to the 3d ADHM quiver theory.

As we have seen earlier, the collective coordinate formalism is a powerful way to relate tRS to $\Delta$BO and eRS to $\Delta$ILW. 
Intuitively, one would expect the $\Delta$ILW to arise as a hydrodynamic limit of eRS, in which the number of particles $n$ is sent to infinity while keeping the density of particles finite. This can be seen from \eqref{keyell} (or its trigonometric version \eqref{keytr}), as this equation implies a relation between eRS and $\Delta$ILW eigenvalues, which greatly simplifies  in the limit $n \rightarrow \infty$ assuming conjecture \eqref{limit} holds. In fact, thanks to the gauge theory computations on both the eRS and $\Delta$ILW sides, we will be able to show explicitly the validity of this conjecture perturbatively in the elliptic deformation $p$. This hints towards an unexpected equivalence at large $n$ between our 5d theory with defects and the 3d ADHM theory.

\subsection{$\Delta$BO from Trigonometric Ruijsenaars-Schneider Model}\label{5.7.1}
Let us first consider the trigonometric case first (see first equation in \eqref{keytr})
\begin{equation}
[\eta(z)]_1 \phi_n(\tau) \vert 0 \rangle = \left[ t^{-n} + t^{-n+1}(1-t^{-1})D^{(1)}_{n,\vec{\tau}}(q,t) \right] \phi_n(\tau) \vert 0 \rangle\,. \label{5.1}
\end{equation}
Here we are taking $|t| >1$; in the opposite case, we need to consider the second equation in \eqref{keytr}. 
We already know that the eigenstates and the eigenvalues of $[\eta(z)]_1$ are labelled by partitions $\lambda$ of $k$ and are independent of length $n$ of the partition. In particular, from \eqref{solBO} we know that the eigenvalues are given by
\begin{equation}
\mathcal{E}_1^{(\lambda)} = 1 - (1-q)(1-t^{-1}) \sum_{(i,j) \in \lambda} q^{i-1}t^{1-j}
= 1 + (1-t^{-1}) \sum_{j=1}^k (q^{\lambda_j}-1)t^{1-j}\,. \label{eigdbo}
\end{equation}
From this expression it is clear that $\lambda_j$ which are equal to zero do not contribute to the final result. On the other hand, the eigenfunctions and the eigenvalues of the tRS model are also labelled by the same partition of $k$, and both of them depend on $n$. Explicitly, the tRS eigenvalues are given by \eqref{tRSev}
\begin{equation}
E_{tRS}^{(\lambda; n)} = \sum_{j=1}^n q^{\lambda_j} t^{n-j}\,.
\end{equation}
Equation \eqref{5.1} is telling us that there is a relation between the $\Delta$BO and tRS eigenvalues: at fixed $\lambda$ we have the following
\begin{equation}
\mathcal{E}_1^{(\lambda)} = t^{-n} + t^{-n+1}(1-t^{-1}) E_{tRS}^{(\lambda; n)} \label{releigtr}
\end{equation}
This equality can be easily shown to be true for all $n$. In fact we can show that 
\begin{equation}
\begin{split}
E_{tRS}^{(\lambda; n)} \,=\, & t^{n-1} \sum_{j=1}^k q^{\lambda_j} t^{1-j} + 
t^{n-1} \sum_{j=1}^n t^{1-j} - t^{n-1} \sum_{j=1}^k t^{1-j} \\
\,=\, & t^{n-1} \sum_{j=1}^k (q^{\lambda_j}-1) t^{1-j} + t^{n-1}\dfrac{1-t^{-n}}{1-t^{-1}}\,,
\end{split}
\end{equation}
which after substitution in \eqref{releigtr} reproduces \eqref{eigdbo}. 

Let us now study what happens in the limit $n \rightarrow \infty$. Even though the limit is not very informative at the trigonometric level, since we already have closed form expressions for the eigenvalues, it will become very important when we discuss the elliptic case. First of all we notice that $\mathcal{E}_1^{(\lambda)}$ and $E_{tRS}^{(\lambda; n)}$ fail to be proportional to each other because of the constant term $t^{-n}$, which however disappears when $n \rightarrow \infty$. This is in agreement with  conjecture \eqref{limit} of \cite{Feigin:2009ab} in the trigonometric limit when the right hand side of \eqref{releigtr} becomes 
\begin{equation}
\lim_{n \rightarrow \infty} \left[ t^{-n+1}(1-t^{-1}) E_{tRS}^{(\lambda; n)} \right] = 
1 + (1-t^{-1})\sum_{j=1}^k (q^{\lambda_j}-1) t^{1-j}\,,
\end{equation}
and coincides with $\mathcal{E}_1^{(\lambda)}$ of \eqref{eigdbo}. 

Therefore, we conclude that there are two ways of recovering the $\Delta$BO eigenvalue from the tRS one at fixed $\lambda$. The first possibility is to use \eqref{releigtr} as it is: the formula works for all $n$, but requires the knowledge of the constant term, which in this case is merely $t^{-n}$. Second, we can take the limit $n \rightarrow \infty$ on the right hand side of \eqref{releigtr}. This method is the mostly relevant if one does not know an explicit expression for the constant term, as it is conjectured to vanish in the limit. Still, one is required to know the eigenvalue for generic $n$. In the following we shall take the large-$n$ limit during the study of the eRS model. 

\subsection{$\Delta$ILW from the Elliptic Ruijsenaars Model}\label{5.7.2}
We have derived the free boson representation of the eRS Hamiltonian in \eqref{keyell}, it reads
\begin{equation*}
\left[ \eta(z;-\tilde{p}q^{-1/2}t^{1/2}) \right]_1 \phi_n(\tau;p) \vert 0 \rangle =  \phi_n(\tau;p)
\left[ t^{-n} \prod_{i=1}^n \dfrac{\Theta_p(qt^{-1}z/\tau_i)}{\Theta_p(qz/\tau_i)} \dfrac{\Theta_p(tz/\tau_i)}{\Theta_p(z/\tau_i)} \eta(z;pq^{-1}t) \right]_1 \vert 0 \rangle
\end{equation*}
\begin{equation}
+ t^{-n+1}(1-t^{-1})\dfrac{(pt^{-1};p)_{\infty}(ptq^{-1};p)_{\infty}}{(p;p)_{\infty}(pq^{-1};p)_{\infty}} D^{(1)}_{n,\vec{\tau}}(q,t;p) \phi_n(\tau;p) \vert 0 \rangle\,.
\label{eq:eRSfreeField}
\end{equation}
We can rewrite the same equality in terms of the eigenvalues
\begin{equation*}
\mathcal{E}_1^{(\lambda)}(\tilde{p}) = \left[ t^{-n} \prod_{i=1}^n \dfrac{\Theta_p(qt^{-1}z/\tau_i)}{\Theta_p(qz/\tau_i)} \dfrac{\Theta_p(tz/\tau_i)}{\Theta_p(z/\tau_i)} \eta(z;pq^{-1}t) \right]_1 
\end{equation*}
\begin{equation}
+ t^{-n+1}(1-t^{-1})\dfrac{(pt^{-1};p)_{\infty}(ptq^{-1};p)_{\infty}}{(p;p)_{\infty}(pq^{-1};p)_{\infty}} E_{eRS}^{(\lambda;n)}(p)\,. \label{releigell}
\end{equation}
Unlike the tRS model, here we no longer know the constant term in the first line on \eqref{releigell} explicitly at finite $n$; therefore, if we want to recover $\mathcal{E}_1^{(\lambda)}(\tilde{p})$ from $E_{eRS}^{(\lambda;n)}(p)$, we should take the large $n$ limit of this equation, which under the conjecture \eqref{limit} reads
\begin{equation}
\mathcal{E}_1^{(\lambda)}(\tilde{p}) = \lim_{n \rightarrow \infty} \left[ t^{-n+1}(1-t^{-1})\dfrac{(pt^{-1};p)_{\infty}(ptq^{-1};p)_{\infty}}{(p;p)_{\infty}(pq^{-1};p)_{\infty}} E_{eRS}^{(\lambda;n)}(p) \right]\,. \label{releigelllim}
\end{equation}
Another problem is that we do not have closed form expressions for the eigenvalues; we can only recover them perturbatively around the trigonometric values, thanks to the computations in gauge theory. In particular, as we have seen that the eigenvalue  $\mathcal{E}_1^{(\lambda)}(\tilde{p})$ for $\Delta$ILW can be obtained from the 3d ADHM theory, with parameters identified as $q=e^{i \gamma \epsilon_1}$, $t=e^{-i \gamma \epsilon_2}$, $\tilde{p}=e^{-2\pi \xi}$, and it is given by \eqref{chern}. On the other hand, the eigenvalue $E_{eRS}^{(\lambda;n)}(p)$ for the eRS model coincides with the Wilson loop VEV \eqref{wilson} for the 5d $\mathcal{N}=1^*$ $U(n)$ theory on $\mathbb{C}^2_{\tilde{\epsilon}_1,\tilde{\epsilon}_2} \times S^1_{\gamma}$ in the NS limit $\tilde{\epsilon}_2 \rightarrow 0$, with Coulomb branch parameters $\mu_a$ fixed by \eqref{mu}; in the latter case $q=e^{i \gamma \tilde{\epsilon}_1}$, $t = e^{-i \gamma m}$ and $p = Q = e^{-8\pi^2 \gamma / g^2_{YM}}$. 

With these results in mind we can verify conjecture \eqref{limit} by proving the validity of \eqref{releigelllim}, in the leading order in the elliptic deformation parameter. Let us demonstrate this for the lowest values of $k$.

\subsubsection{Absence of solitons}
The general strategy is as follows. At fixed $n$, we consider $E_{eRS}^{(\lambda;n)}(p)$ eigenvalue \eqref{wilson} and evaluate it at the values of $\mu_a$ \eqref{mu} corresponding to the partition $\lambda = (0,0,\ldots,0)$ of length $n$. After doing this for the lowest values of $n$, we are able to recognize the dependence on $n$ of the eigenvalue and its behavior at large $n$. 
In the case at hand, this procedure gives us
\begin{equation}
\begin{split}
& t^{-n+1}(1-t^{-1})\dfrac{(pt^{-1};p)_{\infty}(ptq^{-1};p)_{\infty}}{(p;p)_{\infty}(pq^{-1};p)_{\infty}} E_{eRS}^{((0,0,\ldots ,0);n)}(p) = \\ 
& = (1-t^{-n})\left[ 1 + p \dfrac{(1-q)(1-t)(q-t)}{q^2 t (1-q^{-1}t^{1-n})}t^{1-n} + o(p^2) \right]\,,
\end{split}
\end{equation} 
which in the limit $n \rightarrow \infty$ is just $1 + o(p^2)$, in agreement with \eqref{(0)} at order $o(p^2)$.

\subsubsection{One soliton}
Here the relevant partition is $\lambda = (1,0,\ldots,0)$; the eigenvalue depends on $n$ as
\begin{equation}
\begin{split}
& t^{-n+1}(1-t^{-1})\dfrac{(pt^{-1};p)_{\infty}(ptq^{-1};p)_{\infty}}{(p;p)_{\infty}(pq^{-1};p)_{\infty}} E_{eRS}^{((1,0,\ldots ,0);n)}(p) = \\ 
& = \left[ 1-t^{-n} + (q-1)(1-t^{-1}) \right] \\
& - p \dfrac{(1-q)^2(1-t)^2(q-t)(1+q^{-1}t^{1-n})}{q^3(1-q^{-1}t^{2-n})(1-q^{-2}t^{1-n})}t^{-n} \\
& + p \dfrac{(1-q)(1-t)(q-t)(1+q^{-1}t^{1-n})(1-t^{-n})(1 - q^{-2} t^{2-n})}{q t(1-q^{-1}t^{2-n})(1-q^{-2}t^{1-n})} \\
& + o(p^2)\,,
\end{split}
\end{equation} 
which in the limit $n \rightarrow \infty$ reduces to
\begin{equation}
(q + t^{-1} - qt^{-1}) + p \dfrac{(1-q)(1-t)(q-t)}{qt} + o(p^2)\,.
\end{equation}
Comparison with \eqref{(1)} tells us that we have to identify $p = - \tilde{p} \sqrt{qt^{-1}}$ as we anticipated in 
\eqref{anticipation}.

\subsubsection{Two solitons}
\begin{itemize}
\item $k=2$, partition (2,0) \\
For the partition $\lambda = (2,0,\ldots,0)$ we obtain
\begin{equation}
\begin{split}
& t^{-n+1}(1-t^{-1})\dfrac{(pt^{-1};p)_{\infty}(ptq^{-1};p)_{\infty}}{(p;p)_{\infty}(pq^{-1};p)_{\infty}} E_{eRS}^{((2,0,\ldots ,0);n)}(p) = \\ 
& = \left[ 1-t^{-n} + (q^2-1)(1-t^{-1}) \right] \\
& + p \dfrac{(1-q^2)(1-t)^2(q-t)(1-q^{-1}t^{-n})}{t(1-qt)(1-q^{-2}t^{1-n})} \\
& + p \dfrac{(1-q)(1-t)(q-t)(1-q^{-2}t^{-n})(1-q^{-3}t^{2-n})(1-t^{1-n})}{q^2(1-q^{-1}t^{2-n})(1-q^{-2}t^{1-n})(1-q^{-3}t^{1-n})}t^{-n} + o(p^2)\,,
\end{split}
\end{equation}
which in the limit $n \rightarrow \infty$ reduces to
\begin{equation}
(q^2 + t^{-1} - q^2t^{-1}) + p \dfrac{(1-q^2)(1-t)^2(q-t)}{t(1-qt)} + o(p^2)\,.
\end{equation}
This matches \eqref{20} for $p = - \tilde{p} \sqrt{qt^{-1}}$ as expected. 

\item $k=2$, partition (1,1) \\
For the partition $\lambda = (1,1,0,\ldots,0)$ we have
\begin{equation}
\begin{split}
& t^{-n+1}(1-t^{-1})\dfrac{(pt^{-1};p)_{\infty}(ptq^{-1};p)_{\infty}}{(p;p)_{\infty}(pq^{-1};p)_{\infty}} E_{eRS}^{((1,1,0,\ldots ,0);n)}(p) = \\ 
& = \left[ 1-t^{-n} + (q-1)(1-t^{-2}) \right] \\
& + p \dfrac{(1-q)^2(1-t^2)(q-t)(1-t^{1-n})}{qt^2(1-qt)(1-q^{-1}t^{2-n})} \\
& + p \dfrac{(1-q)(1-t)(q-t)(1-q^{-1}t^{-n})(1-q^{-2}t^{3-n})(1-t^{2-n})}{q^2(1-q^{-1}t^{3-n})(1-q^{-1}t^{2-n})(1-q^{-2}t^{1-n})}t^{-n} + o(p^2)\,,
\end{split}
\end{equation}
which in the limit $n \rightarrow \infty$ becomes
\begin{equation}
(q + t^{-2} - qt^{-2}) + p \dfrac{(1-q)^2(1-t^2)(q-t)}{qt^2(1-qt)} + o(p^2)\,.
\end{equation}
This matches \eqref{11} for $p = - \tilde{p} \sqrt{qt^{-1}}$. 

\subsubsection{Three solitons}
\item $k=3$, partition (3,0,0) \\
For the partition $\lambda = (3,0,0,\ldots,0)$ we have
\begin{equation}
\begin{split}
& t^{-n+1}(1-t^{-1})\dfrac{(pt^{-1};p)_{\infty}(ptq^{-1};p)_{\infty}}{(p;p)_{\infty}(pq^{-1};p)_{\infty}} E_{eRS}^{((3,0,0,\ldots ,0);n)}(p) = \\ 
& = \left[ 1-t^{-n} + (q^3-1)(1-t^{-1}) \right] \\
& + p \dfrac{q(1-q^3)(1-t)^2(q-t)(1-q^{-2}t^{-n})}{t(1-q^2t)(1-q^{-3}t^{1-n})} \\
& + p \dfrac{(1-q)(1-t)(q-t)(1-q^{-3}t^{-n})(1-q^{-4}t^{2-n})(1-t^{1-n})}{q^2(1-q^{-1}t^{2-n})(1-q^{-3}t^{1-n})(1-q^{-4}t^{1-n})} t^{-n} + o(p^2)\,,
\end{split}
\end{equation}
which in the limit $n \rightarrow \infty$ becomes
\begin{equation}
\begin{split}
(q^3 + t^{-1} - q^3 t^{-1}) + p \dfrac{q(1-q^3)(1-t)^2(q-t)}{t(1-q^2t)} + o(p^2)\,.
\end{split}
\end{equation}
This matches \eqref{300} for $p = - \tilde{p} \sqrt{qt^{-1}}$. 

\item $k=3$, partition (2,1,0) \\
For the partition $\lambda = (2,1,0,\ldots,0)$ we have
\begin{equation}
\begin{split}
& t^{-n+1}(1-t^{-1})\dfrac{(pt^{-1};p)_{\infty}(ptq^{-1};p)_{\infty}}{(p;p)_{\infty}(pq^{-1};p)_{\infty}} E_{eRS}^{((2,1,0,\ldots ,0);n)}(p) = \\ 
& = \left[ 1-t^{-n} + (q-1)(1-t^{-1})(1+q+t^{-1}) \right] \\
& + p \dfrac{(1-q)(1-t)(q-t)(1-q^2)(1-qt^2)(1-t^{1-n})}{qt^2(1-qt)(1-q^2t)(1-q^{-1}t^{2-n})} \\
& + p \dfrac{(1-q)(1-t)(q-t)(1-t^2)(1-q^2t)(1-q^{-1}t^{-n})}{t(1-qt)(1-qt^2)(1-q^{-2}t^{1-n})} \\
& + p \dfrac{(1-q)(1-t)(q-t)(1-q^{-1}t^{-n+1})(1-q^{-2}t^{-n+3})}{q^2(1-q^{-1}t^{-n+3})(1-q^{-1}t^{-n+2})(1-q^{-2}t^{-n+2})}  \\
& \;\;\;\;\;\;\; \dfrac{(1-q^{-2}t^{-n})(1-q^{-3}t^{-n+2})(1-t^{-n+2})}{(1-q^{-2}t^{-n+1})(1-q^{-3}t^{-n+1})} t^{-n} + o(p^2)\,,
\end{split}
\end{equation}
which in the limit $n \rightarrow \infty$ becomes
\begin{equation}
\begin{split}
& (q^2 + t^{-2} + qt^{-1} - qt^{-2} - q^2 t^{-1}) \\
& + p \dfrac{(1-q)(1-t)(q-t)}{qt^2(1-qt^2)(1-q^2t)} 
\dfrac{[(1-q^2)(1-qt^2)^2 + qt(1-t^2)(1-q^2t)^2]}{(1-qt)} + o(p^2)\,.
\end{split}
\end{equation}
This matches \eqref{210} for $p = - \tilde{p} \sqrt{qt^{-1}}$. 

\item $k=3$, partition (1,1,1) \\
For the partition $\lambda = (1,1,1,0,\ldots,0)$ we have
\begin{equation}
\begin{split}
& t^{-n+1}(1-t^{-1})\dfrac{(pt^{-1};p)_{\infty}(ptq^{-1};p)_{\infty}}{(p;p)_{\infty}(pq^{-1};p)_{\infty}} E_{eRS}^{((1,1,1,0,\ldots ,0);n)}(p) = \\ 
& = \left[ 1-t^{-n} + (q-1)(1-t^{-1})(1+t^{-1}+t^{-2}) \right] \\
& + p \dfrac{(1-q)^2(1-t^3)(q-t)(1-t^{2-n})}{qt^3(1-qt^2)(1-q^{-1}t^{3-n})} \\
& + p \dfrac{(1-q)(1-t)(q-t)(1-q^{-1}t^{-n})(1-q^{-2}t^{4-n})(1-t^{3-n})}{q^2(1-q^{-1}t^{4-n})(1-q^{-1}t^{3-n})(1-q^{-2}t^{1-n})}t^{-n} \\ 
& + o(p^2)\,,
\end{split}
\end{equation}
which in the limit $n \rightarrow \infty$ becomes
\begin{equation}
\begin{split}
(q + t^{-3} - qt^{-3}) + p \dfrac{(1-q)^2(1-t^3)(q-t)}{qt^3(1-qt^2)} + o(p^2)\,.
\end{split}
\end{equation}
This matches \eqref{111} for $p = - \tilde{p} \sqrt{qt^{-1}}$. 
\end{itemize}

\subsection{The Gauge/Hydrodynamics Correspondence} \label{TheGTC}
The above computations suggest the validity of conjecture \eqref{releigelllim}: it is therefore possible to recover the $\Delta$ILW eigenvalues starting from the eRS eigenvalues by taking $n \rightarrow \infty$ limit. This is not surprising from the integrable systems point of view, since $\Delta$ILW is expected to arise as a hydrodynamic limit of eRS; nevertheless, this correspondence looks quite non-trivial from the gauge theory viewpoint in which \eqref{releigelllim} can be rewritten as
\begin{equation}
1 - (1-q)(1-t^{-1})\text{Tr}\, \sigma \big\vert_{\lambda} = \lim_{n \rightarrow \infty} \left[ t^{-n+1}(1-t^{-1}) \left\langle W_{\square}^{U(n)} \right\rangle \right] \Big\vert_{\lambda}\,.
\end{equation}
Here we are proposing an equivalence between an observable in the 3d ADHM theory and a Wilson loop in the 5d $\mathcal{N}=1^*$ $U(n)$ theory at $n \rightarrow \infty$. This indicates an infra-red duality which relates the two theories in the large $n$ limit. In the next section we shall provide further evidence for this correspondence. For clarity let us introduce here the corresponding dictionary:
\begin{center}
\renewcommand\arraystretch{1.2}
\begin{tabular}{|c|c|c|}
\hline
\textbf{elliptic RS} & \textbf{3d ADHM theory} & \textbf{3d/5d coupled theory, $n \rightarrow \infty$} \\ 
\hline coupling $t$ & twisted mass $e^{-i\gamma\epsilon_2}$ & 5d $\mathcal{N}=1^*$ mass deformation $e^{-i\gamma m}$ \\
\hline quantum shift $q$ & twisted mass $e^{i\gamma\epsilon_1}$ & Omega background $e^{i\gamma\widetilde\epsilon_1}$ \\
\hline elliptic parameter $p$ & FI parameter $\tilde{p} = - p/\sqrt{qt^{-1}} $ & 5d instanton parameter $Q$ \\
\hline eigenstates $\lambda$ & ADHM Coulomb vacua & 5d Coulomb branch parameters\\
\hline eigenvalues & $\langle \text{Tr}\,\sigma \rangle$ & $\langle W_{\square}^{U(\infty)} \rangle$ in NS limit $\tilde{\epsilon}_2 \rightarrow 0$\\
\hline
\end{tabular}
\renewcommand\arraystretch{1}
\end{center} \vspace*{0.5 cm}

In general, we expect the ADHM local observable $\langle \text{Tr}\,\sigma^r \rangle$ to be related to the $n \rightarrow \infty$ limit of the 5d Wilson loop $\langle W^{U(n)}_r \rangle$ in the rank $r$ antisymmetric representation. Also note that the second Omega background parameter $\widetilde\epsilon_2$ does not enter the table due to the Nekrasov-Shatashvili limit.

\subsection{Brane Construction}
\label{Sec:StringThDer}
We have demonstrated in this work that the Higgs branch of $U(n)$ 5d $\CN=1^*$ theory in the large-$n$ limit describes the moduli space of $U(1)$ instantons. In this section we shall illustrate this correspondence by using string theory.

First we shall summarize the brane construction of the 5d $\CN=1^*$ theory and describe its Coulomb and Higgs branches along the lines of \cite{Chen:2012we}. The theory in question can be thought of as a lift of the 4d $\CN=2^*$ theory on $\mathbb{R}^4$, whose brane realization was developed in \cite{Witten:1997sc}, to the 5d theory on $\mathbb{R}^4\times S^1$.

\begin{figure}[h]
\centering
\includegraphics[scale=0.52]{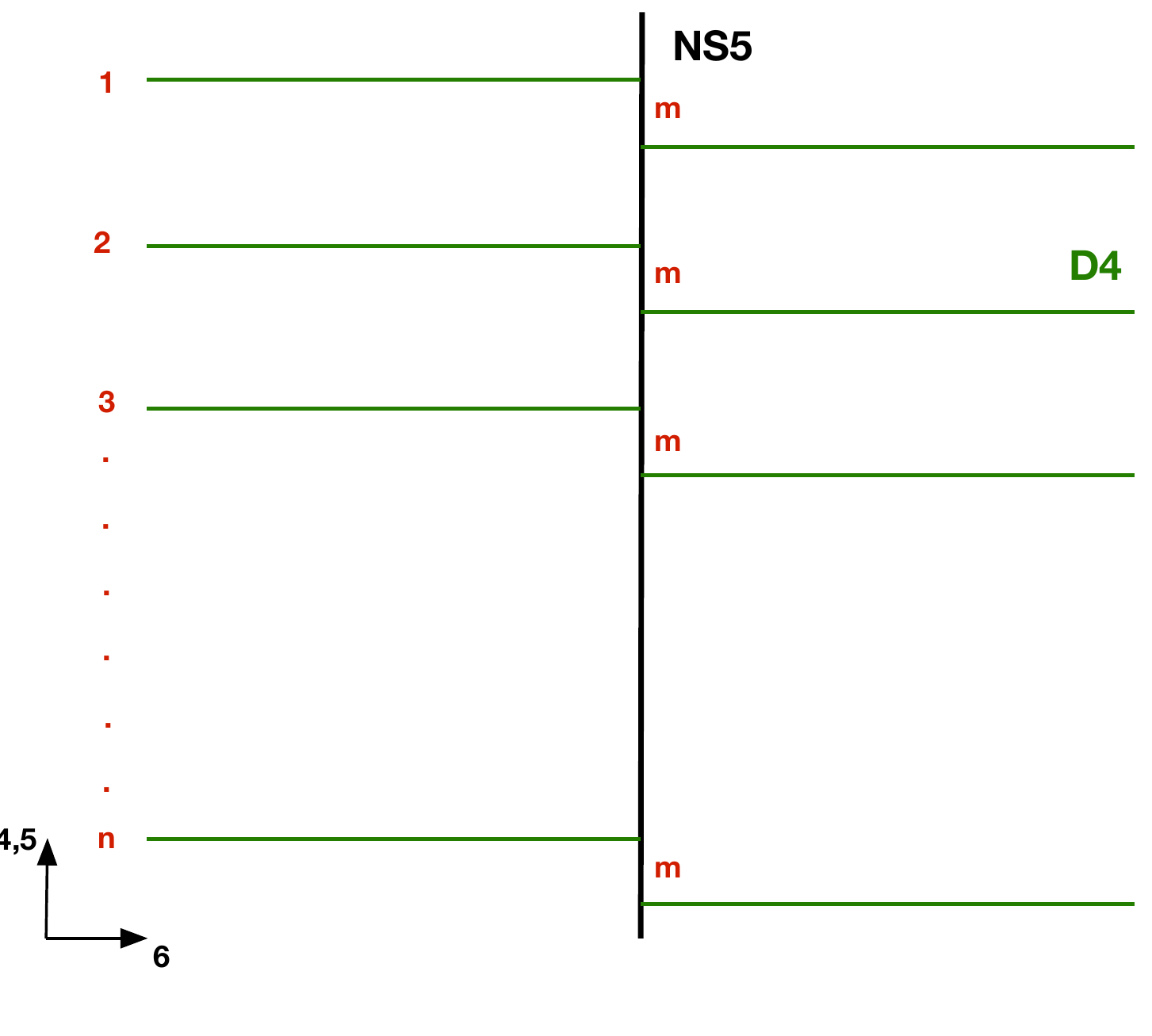} 
\caption{The Coulomb branch of the 5d $\CN=1^*$ $U(n)$ theory realized using Type IIA branes. It is assumed in the picture that the $6$-th direction is periodic and D4 branes reconnect to themselves.} 
\label{fig:n1s5dch}
\end{figure}

The starting point is the Type IIB construction of $U(n)$ maximally supersymmetric Yang-Mills theory in five dimensions which is realized as a theory on $n$ coincident D5 branes along $x^0,x^1,x^2,x^3,x^4,x^6$ with two compact directions -- $x^4$ and $x^6$ with radii $\gamma\gg R_6$ respectively. At the energies much larger than $R_6^{-1}$ we have a 5d theory on $\Reals^4\times S_\gamma^1$. 
For the later purposes it will be convenient to T-dualize along $x^4$ to obtain a Type-IIA description where the D5 branes are located at a point along the $x^4$ circle. Once we turn on the Wilson line for gauge field $A_4$ the stack of branes will in general separate into $n$ branes at different positions along the circle. The positions of branes in 45 plane are given by VEVs of the complex scalar, which, due to the periodicity along $x^4$, is convenient to represent as an exponential $\mu_a = e^{-i\gamma a_a}$, as we used it in \secref{Ruijsenaarssystemsfromgaugetheory}. In order to construct $\CN=1^*$ theory from the $\CN=2$ theory we introduce NS5 brane along 012345 such that the periodic fourbranes break at the position of NS5 brane with an offset in 45 directions given by the mass of the adjoint hypermultiplet $t=e^{-i\gamma m}$, see \figref{fig:n1s5dch}.

In addition we introduce Omega deformation $\epsilon_1$ in 23-plane, which in exponential notation reads $q=e^{i\gamma\epsilon_1}$. As it was discussed in \cite{Chen:2012we} the theory possesses a lattice of Higgs branch loci given in our notation by \eqref{mu} and which is labelled by Young tableaux $\lambda$. In the absence of the Omega deformation all D4 branes merge together consequently into a single spiral-shaped brane which, if the adjoint mass is integral in units of the inversed radius $\gamma^{-1}$, forms a rational winding of the 46-torus. However, when present, the end of $i$th and the beginning of $(i+1)$st 
D4 branes will be offset by $(\lambda_i-\lambda_{i+1})\epsilon_1$. 

\begin{figure}[h]
\centering
\includegraphics[scale=0.45]{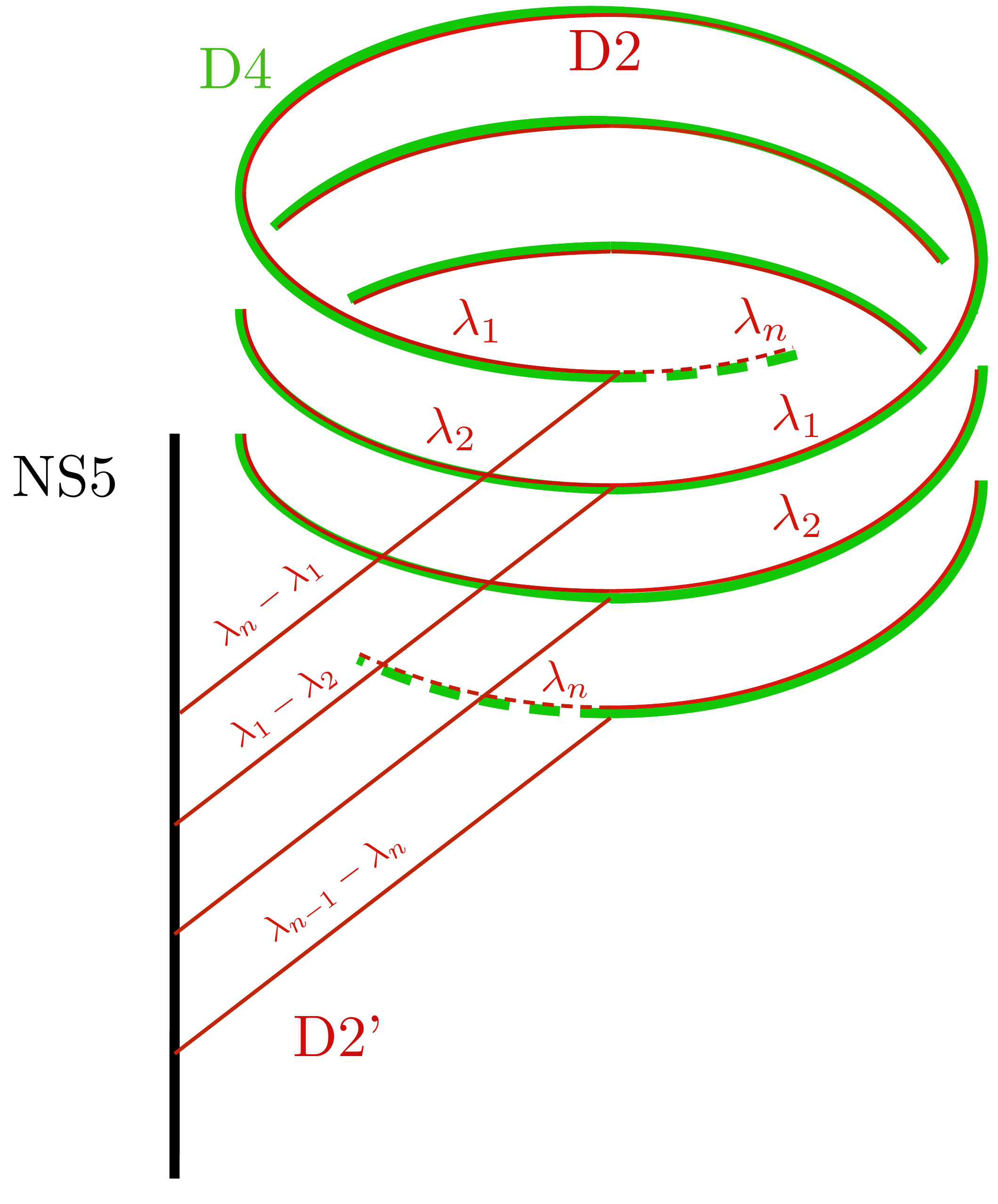} 
\caption{The Higgs branch of the 5d $\CN=1^*$ $U(n)$ theory realized using Type IIA branes. The dashed lines on top and bottom of the picture are identified. Semi-infinite D2' branes are suspended between the NS5 brane and the helical D4 brane which in turn contains circular D2 branes in it.} 
\label{fig:Higgs5d}
\end{figure}

Such brane configuration is known to have a dual description \textit{a l\'a} Gopakumar-Vafa \cite{Gopakumar:1999aa} when
the Coulomb branch is viewed as a resolution of the conifold singularity by a small $S^2$ whose area is proportional to $\epsilon$. The Higgs branch therefore corresponds to the deformation of the conifold  by blowing up a three-sphere, which extends along $x^7$.
This equivalence was used in brane constructions of supersymmetric gauge theories in Omega background abundantly, see e.g. \cite{Dorey:2011pa, Chen:2011sj, Bulycheva:2012aa, Aganagic:2013aa, Aganagic:2014aa}. In other words, one needs to accommodate extra units of magnetic flux through 23-plane for each Cartan direction of the gauge group given by $i$th column of $\lambda$. As was noted in \cite{Chen:2012we} this can be achieved by allocating the corresponding number of D2 branes along 01 plane and which wrap a single direction in 46-torus, which needs to be done supersymmetrically. Since $i$th D4 brane worldvolume should now contain $\lambda_i$ units of magnetic flux there is a jump of the number of D2 branes for $i$th and $(i+1)$st D4 branes at the location of the NS5 brane where they meet. Since the orientation of these fourbranes is mutually opposite there is an excess of $\lambda_i-\lambda_{i+1}$ D2 brane charge, which should be compensated by adding additional D2 branes stretching from the D4 branes to the NS5 brane, see \figref{fig:Higgs5d}. More precisely there are now two types of D2 branes -- first branes that wrap 01-plane and a 1-cycle in $T^2_{46}$ and those along 017 directions. 
The former will be called D2' and the mutual orientation of all branes in this phase is shown in the table below.
\begin{center}
\begin{tabular}{|c|c|c|c|c|c|c|c|c|c|c|}
\hline
         & 0 & 1 &  2 & 3 & 4 & 5 & 6 & 7 & 8 & 9 \\
\hline \hline
NS5 & x &  x &  x &  x & x  & x &    &       &   &   \\
\hline
D4    & x &  x & x  & x  &  $\cos\delta$ &   & $\sin\delta$ &    &     &    \\
\hline
D2'    & x &  x &  &  &    &     &    &  x &      &    \\
\hline
D2    & x &  x &  &  & $-\sin\delta$   &     &  $\cos\delta$  &   &      &    \\
\hline
\end{tabular}\label{tab:HWbranes}
\end{center}
Here $\delta\sim \tan^{-1}m/R$ is winding angle of branes on the fundamental domain of $T^2_{46}$ which can be represented as a square in 46-plane. We can see that D2s and D4s form two orthogonal rational windings of the torus.

One immediately notices that the net charge of all D2' branes is zero, the fact which we shall use momentarily as we explore the large-$n$ limit. At the Higgs branch locus \eqref{mu} we can compute spectra of both elliptic Ruijsenaars-Schneider and $\Delta$ILW models and compare them \eqref{eq:eRSfreeField}. 
We can see that for finite $n$ there are two contributions to the energy with the additional term present in the first line of \eqref{eq:eRSfreeField}, which is suppressed by $t^{-n}$. 

Thus what happens with the brane system shown in \figref{fig:Higgs5d} as we send $n$ to infinity? We shall apply a certain scaling to the vertical direction in the figure such that the size of the D4 brane helix remains fixed, but the number of fourbranes goes to infinity. This can be done, for example, if we scale the $\gamma$ with $n$ appropriately. The number of circular D2 branes is finite and is given by the number of columns of $\lambda$. Therefore in the scaled picture, where the size of the vertical direction (which is periodic) in \figref{fig:Higgs5d} is $2\pi \gamma$, all D2 branes will be located on the top, or at a point in $x^4$. The total number of D2 branes is $\sum_i\lambda_i=k$, whereas, as we have already noticed above, all D2's disappear and the NS5 brane completely detaches from the system! We can now see that only a single long D4 brane remains together with $k$ D2 branes. Note that the scaling limit which we have taken made winding angle $\delta$ to be effectively zero, so D2 branes now wrap 016 directions. 

Finally, in order to recover the desired ADHM brane construction, we T-dualize our setup along $x^6$ to obtain D1/D5 system. One can additionally apply another T-duality along a circle inside 6789, which, if we recall the dual description of the NS5 brane, represents the Taub-NUT circle. The last T-duality brings us to $k$ D2 branes probing a single D6 brane.
\begin{center}
\begin{tabular}{|c|c|c|c|c|c|c|c|c|c|c|}
\hline
         & 0 & 1 &  2 & 3 & 4 & 5 & 6 & 7 & 8 & 9 \\
\hline \hline
D6    & x &  x & x  & x  & x  &   &x  &    & x    &    \\
\hline
D2    & x &  x &  &  &    &     &    &   &   x   &    \\
\hline
\end{tabular}\label{tab:HWbranes}
\end{center}
This concludes our derivation of the eRS/$\Delta$ILW correspondence detailed in \secref{TheGTC}.

\section{Future Directions}
\label{Sec:Future}
Clearly much more can be said about dualities between supersymmetric gauge theories and integrable many-body and hydrodynamical systems. Below we list, in a random order, some of questions which need to be answered in the future publications.

\begin{itemize}

\item Our analysis was restricted to Abelian ILW$_1$ systems. One needs to extend the eRS/ILW correspondence to multi-dimensional ILW$_N$ systems. Our expectations suggest that one needs to study 5d theories on ALE singularities and then take large-n limit, which in the end should lead to the $N$-instanton ADHM quiver \cite{2014JHEP...07..141B,2015arXiv150507116B}.

\item In the paper we discussed the ILW system and its BO limit $\delta \rightarrow \infty$, but we did not consider the KdV limit $\delta \rightarrow 0$ (see \secref{Sec:ILWDescription}). Naively the KdV limit seems to be ill-defined: for example, the ADHM Bethe Ansatz Equations appear to be incomplete for $\delta \rightarrow 0$. This might be related to the fact that in order to recover the KdV equation from the ILW equation a shift $u \rightarrow u + \delta^{-1}$ of the wave profile is required, and this makes the profile singular at $\delta = 0$. From the mathematical point of view this singularity should correspond to the singular point of the ADHM quantum cohomology, which happens presicely at $\delta = 0$, where $\delta$ is interpreted as a K\"{a}hler modulus of the instanton moduli space $\mathcal{M}_{k,1}$. One may try to avoid the problem by considering a complexified K\"{a}hler modulus with a non-zero $\theta$-angle in the ADHM theory, however the complex nature of this parameter is lacking a physics interpretation in terms of the hydrodynamic system. Our feeling is that, as suggested in \cite{2013JHEP...11..155L}, the correct way of studying the KdV limit is to consider the $\text{ILW}_2$ system which, for $\delta \rightarrow 0$, should decompose into a free field and another field satisfying the KdV equation. Further investigation on this point is needed.

\item As we have seen in the main text, the $\Delta$BO$_1$ system reduces to the BO$_1$ system when $\gamma \rightarrow 0$, i.e. $q,t \sim 1 + \ldots$. On the other side of the correspondence the trigonometric Ruijsenaars-Schneider model reduces to the trigonometric Calogero model (Macdonald polynomials become Jack polynomials). One could then study Macdonald polynomials at $p$-th root of unity (the so-called Uglov polynomials \cite{1998CMaPh.191..663U}), which were shown to appear in the instanton counting on ALE spaces $\mathbb{C}^2/\mathbb{Z}_p$ in \cite{Belavin:2012eg} and also in generalized BO$_{1,p}$ systems in \cite{2015JHEP...02..150A} (see also \cite{2015arXiv150507116B}). It would be interesting to verify that our $\Delta$BO$_1$ and $\Delta$ILW$_1$ systems reduce to these generalized BO$_{1,p}$ and ILW$_{1,p}$ systems at in the root of unity.

\item The mathematical meaning of polynomial ring \eqref{eq:eRSRing} needs to be properly understood. We hope that our observation regarding its relationship with quantum K-theory of the ADHM quiver may help to better understand this object.

\item In this paper we have studied large-$n$ limit of elliptic Ruijsenaars-Schneider model coming from 5d $\CN=1^*$ gauge theory compactified on a circle. The next natural generalization of our construction would be to study torus compactifications of 6d theories and their large-$n$ physics.

\item In \cite{Buryak:2015uq,Buryak:2014fk} the authors gave quite a generic quantization scheme of many known integrable systems of evolutionary PDEs using methods of intersection theory of the moduli space of curves. The generating function of integrable Hamiltonians that we have used in the paper also appears in their context. It is worthwhile realizing a deeper connection between the two approaches.

\end{itemize}

\section*{Acknowledgements}
We would like to thank Sergei Gukov, Davide Gaiotto, Mina Aganagic, Shamil Shakirov, Nathan Haouzi, Alessandro Tanzini, Giulio Bonelli, Hee-Cheol Kim, Benjamin Assel, Stefano Cremonesi, Cristian Vergu for fruitful and insightful discussions.

This work was performed in part at the Aspen Center for Physics, which is supported by National Science Foundation grant PHY-1066293.
PK thanks W. Fine Institute for Theoretical Physics at University of Minnesota, Kavli Institute for Theoretical Physics at University of California Santa Barbara, University of California at Berkeley, International School of Advanced Studies (SISSA) in Trieste, and Simons Center for Geometry and Physics where part of his work was done, for kind hospitality. 
AS thanks the Perimeter Institute for Theoretical Physics and the Department of Mathematics at King's College London for their hospitality during the completion of this work. 

Our research was partly supported by the Perimeter Institute for Theoretical Physics. Research at Perimeter Institute is supported by the Government of Canada through Industry Canada and by the Province of Ontario through the Ministry of Economic Development and Innovation. The work of AS is supported in part by an Erasmus Placement Fellowship and by the INFN project ST$\&$FI.

\newpage

\appendix

\section{The ADHM quiver and Bethe Ansatz Equations for ILW} \label{appA}

In this Appendix we will consider the $\mathcal{N} = 2^*$ ADHM quiver theory on $\Complex \times S^1_{\gamma}$ (or $\mathbb{P}^1 \times S^1_{\gamma}$) inside the 11d geometry $\Complex_q \times \Complex_t \times \Complex \times \mathcal{O}(-2)_{\mathbb{P}^1} \times S^1_{\gamma}$. The field content of the quiver is summarized in the table below.

\begin{table}[h!]
\begin{center}
\begin{tabular}{c|c|c|c|c|c}
{} & $\chi$ & $B_{1}$ & $B_{2}$ & $I$ & $J$ \\ \hline
D-brane sector & D2/D2 & D2/D2 & D2/D2 & D2/D6 & D6/D2 \\ \hline
gauge $U(k)$ & $Adj$ & $Adj$ & $Adj$ & $\mathbf{k}$ & $\mathbf{\bar{k}}$ \\ \hline
flavor $U(N)\times U(1)^{2}$ & $\mathbf{1}_{(-1,-1)}$ & $\mathbf{1}_{(1,0)}$ & $\mathbf{1}_{(0,1)}$ & $\mathbf{\bar{N}}_{(0,0)}$ & $\mathbf{N}_{(1,1)}$ \\ \hline
twisted masses & $\epsilon_1 + \epsilon_2$ & $-\epsilon_{1}$ & $-\epsilon_{2}$ & $-a_{j}$ & $a_{j}-\epsilon_1 - \epsilon_2$ \\ \hline
$R$-charge & $2$ & $0$ & $0$ & $0$ & $0$ \\ \hline
\end{tabular} 
\caption{Matter content of the ADHM Gauged Linear Sigma Model.}
\end{center}
\end{table} 

\begin{figure}[h!]
  \centering
\includegraphics[width=0.6\linewidth]{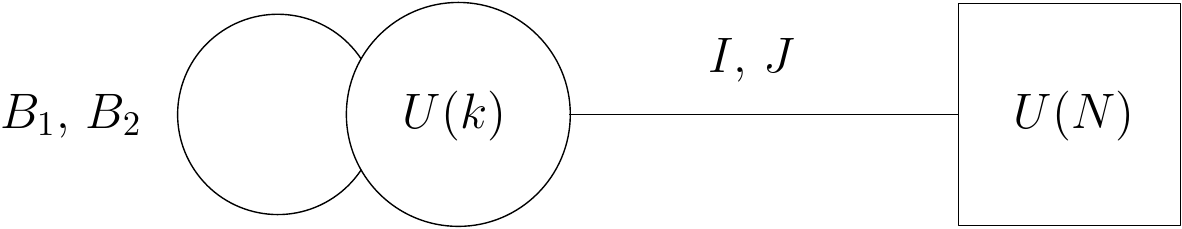} 
  \caption{The ADHM quiver.\label{fig:A}}
\end{figure}

\noindent The superpotential of the theory is given by $W=\textrm{Tr}_{k}\left\{\chi\left([B_{1},B_{2}]+IJ\right)\right\}$.
In the case $\epsilon_1 + \epsilon_2 = 0$ the $\mathcal{N}=2$ vector supermultiplet and the $\mathcal{N}=2$ adjoint chiral supermultiplet $\chi$ combine into an $\mathcal{N}=4$ vector supermultiplet; on the other hand, when $\epsilon_1 + \epsilon_2 \neq 0$ supersymmetry is broken to $\mathcal{N} = 2^*$.
The moduli space $\mathcal{M}_{k,N}$ of supersymmetric vacua in the Higgs branch is obtained by setting to zero the VEV of the adjoint scalar field in the $\chi$ supermultiplet and it is given by the solutions of the $F$ and $D-$term equations, modulo the action of the gauge group $U(k)$. More explicitly we have \vspace*{0.3 cm}
\begin{center}
\begin{tabular}{l}
$\mathcal{M}_{k,N}$ = \Bigg\{ 
\begin{tabular}{ll}
$[B_{1},B_{2}] + IJ = 0$ & ($F$-term) \\ 
$[B_{1},B_{1}^{\dagger}] + [B_{2},B_{2}^{\dagger}] + I I^{\dagger} -J^{\dagger}J = \xi$ & ($D$-term) \\ 
\end{tabular} 
\Bigg\} \Bigg/ $U(k)$\,,
\end{tabular} 
\end{center} \vspace*{0.3 cm}
where $\xi$ is Fayet-Iliopoulos parameter. This manifold can be immediately identified with the ADHM moduli space of $k$ instantons for a pure $U(N)$ Yang-Mills theory. In fact, thinking in terms of a $D2/D6$ brane system, the $k$ $D2$ branes wrapped on $\mathbb{P}^1 \times S^1_{\gamma}$ can be understood as a $k$-instanton for the pure $U(N)$ supersymmetric theory living on the $N$ $D6$ branes which wrap $\Complex_q \times \Complex_t \times \mathbb{P}^1 \times S^1_{\gamma}$ (here $q = e^{i \gamma \epsilon_1}$, $t = e^{-i \gamma \epsilon_2}$). As it is well known in the context of $D(p-4)/Dp$ brane systems, the auxiliary 3d theory living on the $D2$ branes is precisely our ADHM quiver theory and describes the instanton moduli space $\mathcal{M}_{k,N}$. When the radius of the $S^1_{\gamma}$ circle is sent to zero we go back to the setting of \cite{2014JHEP...07..141B} with a system of $k$ $D1$ and $N$ $D5$ branes wrapping respectively $\mathbb{P}^1$ and $\Complex_q \times \Complex_t \times \mathbb{P}^1 $ inside the 10d geometry $\Complex_q \times \Complex_t \times \Complex \times \mathcal{O}(-2)_{\mathbb{P}^1}$.
 
One knows from the Bethe/Gauge correspondence that the equations determining the Coulomb branch vacua coincide with Bethe Ansatz Equations for a quantum integrable system; in this case, following \cite{Nekrasov:2009ui, Nekrasov:2009uh} we obtain (for $N=1$ with relative to $k$ ``instantons" for the 7d $U(1)$ gauge theory living on the $D6$ brane)
\begin{equation}
\begin{split}
& 
\sin [\frac{\gamma}{2} (\Sigma_s - a)]
\prod_{\substack{t = 1 \\ t\neq s}}^k \dfrac{\sin [\frac{\gamma}{2} (\Sigma_{st} - \epsilon_1)] \sin [\frac{\gamma}{2} (\Sigma_{st} - \epsilon_2)]}{\sin [\frac{\gamma}{2} ( \Sigma_{st})] \sin [\frac{\gamma}{2} (\Sigma_{st} - \epsilon)]} = \\
& e^{-2 \pi \xi}\, \sin [\frac{\gamma}{2} ( -\Sigma_s + a - \epsilon)] 
\prod_{\substack{t = 1 \\ t\neq s}}^k \dfrac{\sin [\frac{\gamma}{2} ( \Sigma_{st} + \epsilon_1)] \sin [\frac{\gamma}{2} (\Sigma_{st} + \epsilon_2)]}{\sin [\frac{\gamma}{2} ( \Sigma_{st})] \sin [\frac{\gamma}{2} (\Sigma_{st} + \epsilon)]} \label{BAEapp}
\end{split}
\end{equation}
We can as well set $a=0$ in the following. Here $\epsilon = \epsilon_1 + \epsilon_2$ and $\Sigma_s$ are the scalars in the 2d $\mathcal{N} = (2,2)$ superfield strength multiplet arising when $S^1_{\gamma}$ shrinks to zero size; the effect of the finite-size $S^1_{\gamma}$ circle consists in having to take into account all the Kaluza-Klein modes, which generate the sine functions in \eqref{BAEapp}.

When $\xi \rightarrow \infty$, the solutions to \eqref{BAEapp} are labelled by partitions $\lambda$ of $k$, and are given by 
\begin{equation}
\Sigma_s = (i-1)\e_1 + (j-1)\e_2 \;\;\;\text{mod } 2 \pi i\,.
\end{equation} 
For $\xi$ finite we can define $\sigma_s = e^{i \gamma \Sigma_s}$, $q = e^{i \gamma \epsilon_1}$, $t = e^{-i \gamma \epsilon_2}$ and rewrite \eqref{BAEapp} as 
\begin{equation}
\begin{split}
& (\sigma_s - 1) \prod_{\substack{t = 1 \\ t\neq s}}^k \dfrac{(\sigma_{s} - q \sigma_t) (\sigma_{s} - t^{-1} \sigma_t)}{( \sigma_{s} - \sigma_t) (\sigma_{s} - q t^{-1} \sigma_{t})} =
-e^{-2\pi\xi}\sqrt{q t^{-1}} \, (\sigma_s - q^{-1}t) \prod_{\substack{t = 1 \\ t\neq s}}^k \dfrac{(\sigma_{s} - q^{-1} \sigma_t) (\sigma_{s} - t \sigma_t)}{(\sigma_{s} - \sigma_t) (\sigma_{s} - q^{-1} t \sigma_{t})}\,. \label{BAE2app}
\end{split}
\end{equation}
Our result states that these equations coincide with the Bethe Ansatz Equations for quantum $\Delta$ILW model. Perturbatively in $e^{-2\pi\xi}$ the solutions to \eqref{BAE2app} are still labelled by partitions $\lambda$ of $k$, and the eigenvalue of the first $\Delta$ILW Hamiltonian is given by
\begin{equation}
\mathcal{E}_{1}^{(\lambda)} = 1 - (1-q)(1-t^{-1}) \,\text{Tr}\sigma \big\vert_{\lambda} 
\end{equation} 
where $\text{Tr}\sigma = \sum_s \sigma_s = \sum_s e^{i \gamma \Sigma_s}$ is evaluated at the solutions of \eqref{BAE2app}.

\section{Comments on the reduction to ILW} \label{appB}

In \cite{Feigin:2009ab} the authors noticed that the $\gamma$ expansion of $\eta_0$ can be related to the operator of quantum multiplication in the small quantum cohomology ring of the instanton moduli space $\mathcal{M}_{k,1}$ introduced in  \cite{2004math.....11210O}, which as discussed in \cite{2014JHEP...07..141B} coincides with the quantum ILW Hamiltonian $\widehat{I}_3$. In particular, we have
\begin{equation}
\eta_0 = [\eta(z;-\tilde{p}q^{-1/2}t^{1/2})]_1 = 1 + \gamma^2 \widehat{I}_2 + \gamma^3 \widehat{I}_3 + \gamma^4 \widehat{I}_4 + \ldots \label{gf}
\end{equation}
Since $\Delta$ILW reduces to ILW as $\gamma \rightarrow 0$, given that the time evolution for quantum $\Delta$ILW is given by $\widehat{\mathcal{H}}_1 = \eta_0$, and taking into account that in the limit $qt^{-1} = 1$ (i.e. $\epsilon_1 + \epsilon_2 = 0$) equation \eqref{ILWeqqu} reduces to the Hopf (dispersionless KdV) equation and \eqref{gf} reduces to the generating function for the quantum Hopf Hamiltonians studied in \cite{0036-0279-58-5-R03}\footnote{The same generating function appears in \cite{Eli,2008JGP....58..931R,2014arXiv1407.5824D} in relation to Symplectic Field Theory.}, one could expect $\eta_0$ to be a \textit{generating function} for the ILW quantum Hamiltonians $\widehat{I}_l$. Nevertheless, this turns out not to be true: (\textbf{EXPLANATION})\footnote{We thank Paolo Rossi for this argument}. 
 
In any case, the first Hamiltonians $\widehat{I}_2$, $\widehat{I}_3$ and $\widehat{I}_4$ obtained from \eqref{gf} belong to the set of quantum ILW Hamiltonians: in this Appendix we will study them and their eigenvalue equation, and show that the results are indeed in agreement with the $\gamma$ expansion of the corresponding $\Delta$ILW eigenvalue $\mathcal{E}_1$.
 
Again, we start from
\begin{equation}
\eta(z;pq^{-1}t) = \text{exp}\left(\sum_{n>0}\lambda_{-n}z^n\right) \text{exp}\left(\sum_{n>0}\lambda_{n}z^{-n}\right)\,, 
\end{equation}
with commutation relations for the $\lambda_m$
\begin{equation}
[\lambda_m, \lambda_n] = -\dfrac{1}{m} \dfrac{(1-q^m)(1-t^{-m})(1-(pq^{-1}t)^m)}{1-p^m} \delta_{m+n,0}\,.
\end{equation}
It appears to be more convenient to use the following normalization for the oscillators
\begin{equation}
\lambda_m = \dfrac{1}{\vert m \vert} \sqrt{-\dfrac{(1-q^{\vert m \vert})(1-t^{-\vert m \vert})(1-(pq^{-1}t)^{\vert m \vert})}{1-p^{\vert m \vert}}} \overline{a}_m\,,
\end{equation}
with commutation relations
\begin{equation}
[\overline{a}_m, \overline{a}_n] = m \delta_{m+n,0}\,.
\end{equation}
After substituting $p = -\tilde{p} \sqrt{qt^{-1}}$ we arrive at
\begin{equation}
\begin{split}
\lambda_m \;=\; & \dfrac{1}{\vert m \vert} \sqrt{-\dfrac{(1-q^{\vert m \vert})(1-t^{-\vert m \vert})(1-(-\tilde{p}q^{-1/2}t^{1/2})^{\vert m \vert})}{1-(-\tilde{p}q^{1/2}t^{-1/2})^{\vert m \vert}}} \overline{a}_m \;= \\
=\; & \gamma \sqrt{\e_1 \e_2} \Bigg[ 1 + i \gamma \dfrac{\e_1 + \e_2}{4} m \dfrac{1 + (-\tilde{p})^m}{1 - (-\tilde{p})^m} \;+ \\
& + \gamma^2 \left( - \dfrac{(\e_1 + \e_2)^2}{8} m^2 \dfrac{(-\tilde{p})^m}{(1 - (-\tilde{p})^m)^2} - m^2 \dfrac{5 (\e_1 + \e_2)^2 - 4 \e_1 \e_2}{96} \right) + \ldots \Bigg] \overline{a}_m \,.
\end{split}
\end{equation}
Next we expand $\eta_0$ in powers of $\gamma$ as in \eqref{gf} and get
\begin{equation} 
\widehat{I}_2 = \e_1 \e_2 \sum_{m>0} \overline{a}_{-m}\overline{a}_m\,, 
\end{equation}
\begin{equation}
\widehat{I}_3 = i \e_1 \e_2  \dfrac{\epsilon_1 + \epsilon_2}{2} \sum_{m>0} m \dfrac{1 + (-\tilde{p})^m}{1 - (-\tilde{p})^m} \, \overline{a}_{-m} \overline{a}_m + \dfrac{(\e_1 \e_2 )^{\frac{3}{2}}}{2} \sum_{m,n > 0} (\overline{a}_{-m-n} \overline{a}_m \overline{a}_n + \overline{a}_{-m} \overline{a}_{-n} \overline{a}_{m+n})\,, 
\end{equation}
and
\begin{equation*}
\widehat{I}_4 = \dfrac{(\e_1 \e_2)^2}{6} \sum_{m,n,l > 0} (\overline{a}_{-m-n-l} \overline{a}_m \overline{a}_n \overline{a}_l + \overline{a}_{-m}\overline{a}_{-n}\overline{a}_{-l}\overline{a}_{m+n+l})
+ \dfrac{(\e_1 \e_2)^2}{4} \sum_{\substack{m,n,l,k > 0 \\ m+n=l+k}} \overline{a}_{-m}\overline{a}_{-n} \overline{a}_l \overline{a}_k
\end{equation*}
\begin{equation*}
\begin{split}
+ i (\e_1 \e_2)^{\frac{3}{2}} \dfrac{\e_1 + \e_2}{8} \sum_{m,n>0} & \Big[ m \dfrac{1 + (-\tilde{p})^m}{1 - (-\tilde{p})^m} + n \dfrac{1 + (-\tilde{p})^n}{1 - (-\tilde{p})^n} \\
& + (m+n) \dfrac{1 + (-\tilde{p})^{m+n}}{1 - (-\tilde{p})^{m+n}} \Big](\overline{a}_{-m-n} \overline{a}_m \overline{a}_n + \overline{a}_{-m} \overline{a}_{-n} \overline{a}_{m+n})
\end{split}
\end{equation*}
\begin{equation}
- \e_1 \e_2 \dfrac{2(\e_1 + \e_2)^2 - \epsilon_1 \epsilon_2}{12} \sum_{m>0} m^2 \overline{a}_{-m}\overline{a}_m - \e_1 \e_2 \dfrac{(\e_1 + \e_2)^2}{2} \sum_{m>0} m^2 \dfrac{(-\tilde{p})^m}{(1-(-\tilde{p})^m)^2} \overline{a}_{-m}\overline{a}_m\,,
\end{equation}
which coincide with the first known ILW Hamiltonians.

\subsection{The ILW Spectrum}

Let us now consider the eigenvalue problem for these quantum Hamiltonians, as we did in Section \ref{Sec:somesec}. Denoting by $k$ the eigenvalue of $\widehat{I}_2$, in other words, the number of solitons present in the ILW solution, we restrict ourselves to the cases with $k=2$ and $k=3$ in the following. 

\subsubsection{Two soliton configuration} 
A state with $k=2$ can generically be written as
\begin{equation}
(c_1 \overline{a}_{-1}^2 + c_2 \overline{a}_{-2}) \vert 0 \rangle
\end{equation}
in terms of two constants $c_1$, $c_2$ which are to be determined. The eigenvalue equation for the $\widehat{I}_3$ Hamiltonian 
\begin{equation*}
\widehat{I}_3 (c_1 \overline{a}_{-1}^{2} + c_2 \overline{a}_{-2}) \vert 0 \rangle \;=\; E_3 (c_1 \overline{a}_{-1}^2 + c_2 \overline{a}_{-2}) \vert 0 \rangle =
\end{equation*}
\begin{equation}
= \left[ \left((\e_1 \e_2)^{3/2} c_2 + i \e_1 \e_2(\e_1 + \e_2) \dfrac{1 - \tilde{p}}{1 + \tilde{p}} c_1 \right) \overline{a}_{-1}^2 
+ \left((\e_1 \e_2)^{3/2} c_1 + 2 i \e_1 \e_2(\e_1 + \e_2) \dfrac{1 + \tilde{p}^2}{1 - \tilde{p}^2} c_2 \right) \overline{a}_{-2} \right] \vert 0 \rangle
\end{equation}
results in the following equation for the energy
\begin{equation}
\left(E_3 - i \e_1 \e_2 (\e_1 + \e_2) \dfrac{1 - \tilde{p}}{1 + \tilde{p}}\right) \left(E_3 - 2 i \e_1 \e_2 (\e_1 + \e_2) \dfrac{1 + \tilde{p}^2}{1 - \tilde{p}^2}\right) = (\e_1 \e_2)^{3}\,,
\end{equation}
which has the two solutions
\begin{equation}
\begin{split}
& \frac{E_3^{(1)}}{\e_1 \e_2} = i (2\e_1 + \e_2) + \tilde{p} \dfrac{2 i (\e_1 + \e_2) \e_2}{\e_1 - \e_2} + \tilde{p}^2 \dfrac{2 i (\e_1 + \e_2) (2 \e_1^3 - 7 \e_1^2 \e_2 + 2 \e_1 \e_2^2 - \e_2^3)}{(\e_1 - \e_2)^3} 
+ o(\tilde{p}^3)\,, \\
& \frac{E_3^{(2)}}{\e_1 \e_2} = i (\e_1 + 2\e_2) + \tilde{p} \dfrac{2 i (\e_1 + \e_2) \e_1}{\e_2 - \e_1} + \tilde{p}^2 \dfrac{2 i (\e_1 + \e_2) (2 \e_2^3 - 7 \e_2^2 \e_1 + 2 \e_2 \e_1^2 - \e_1^{3})}{(\e_2 - \e_1)^3}  
+ o(\tilde{p}^3)\,. \label{k2E3}
\end{split} 
\end{equation}
Similarly, the eigenvalue equation for $\widehat{I}_4$
\begin{equation}
\widehat{I}_4 (c_1 \overline{a}_{-1}^{2} + c_2 \overline{a}_{-2}) \vert 0 \rangle 
= E_4 (c_1 \overline{a}_{-1}^2 + c_2 \overline{a}_{-2}) \vert 0 \rangle
\end{equation}
results in equation
\begin{equation}
\begin{split}
& \left[E_4 + \e_1 \e_2(\e_1 + \e_2)^2 \left( \dfrac{1}{3} + \dfrac{\tilde{p}^2}{(1-\tilde{p}^2)^2} \right) - \dfrac{2 (\e_1 \e_2)^2}{3}\right]
\left[E_4 + 4 \e_1 \e_2(\e_1 + \e_2)^2 \left( \dfrac{1}{3} + \dfrac{\tilde{p}^2}{(1-\tilde{p}^2)^2} \right) - \dfrac{2 (\e_1 \e_2)^2}{3}\right] \\
& = - \dfrac{(\e_1 \e_2)^3(\e_1 + \e_2)^2}{4} \left( \dfrac{1 - \tilde{p}}{1 + \tilde{p}} + \dfrac{1 + \tilde{p}^2}{1 - \tilde{p}^2} \right)^2 
\end{split}
\end{equation}
with solutions
\begin{equation*}
\begin{split}
\frac{E_4^{(1)}}{\e_1 \e_2} =&  - \left( \dfrac{\e_2^{2}}{3} + \e_1 \e_2 + \dfrac{4 \e_1^2}{3} \right) - \tilde{p} \dfrac{(\e_1 + \e_2) \e_2 (3 \e_1 + \e_2)}{\e_1 - \e_2} \\
& + \tilde{p}^2 \dfrac{2 (\e_1 + \e_2) (- 2 \e_1^4 + 7 \e_1^3 \e_2 + \e_1^2 \e_2^2 + \e_1 \e_2^3 + \e_2^4)}{(\e_1 - \e_2)^3} 
+ o(\tilde{p}^3) 
\end{split} 
\end{equation*}
\begin{equation}
\begin{split}
\frac{E_4^{(2)}}{\e_1 \e_2} =& - \left( \dfrac{\e_1^{2}}{3} + \e_1 \e_2 + \dfrac{4 \e_2^{2}}{3} \right) - \tilde{p} \dfrac{(\e_1 + \e_2) \e_1 (3 \e_2 + \e_1)}{\e_2 - \e_1} \\
& + \tilde{p}^2 \dfrac{2 (\e_1 + \e_2) (-2 \e_2^{4} + 7 \e_2^3 \e_1 + \e_1^2 \e_2^2 + \e_2 \e_1^3 + \e_1^4)}{(\e_2 - \e_1)^3}  
+ o(\tilde{p}^3) \label{k2E4}
\end{split} 
\end{equation}
It is easy to check that \eqref{k2E3}, \eqref{k2E4} coincide with the $\gamma$ expansions at orders $\gamma^3$, $\gamma^4$ of \eqref{20osc} and \eqref{11osc}.

As a final comment, let us notice here that in the Benjamin-Ono limit $\tilde{p} \rightarrow 0$ the eigenstates become
\begin{equation}
\begin{split}
& (\overline{a}_{-1}^2 + i \e_1 \overline{a}_2 ) \vert 0 \rangle\,, \\
& (\overline{a}_{-1}^2 + i \e_2 \overline{a}_2 ) \vert 0 \rangle\,. \label{es2}
\end{split} 
\end{equation}
In the spirit of isomorphism \eqref{iso}, the above states can be compared with the $\gamma \rightarrow 0$ limit of the Macdonald polynomials of \eqref{mac}, given by Jack polynomials $p_1^2 - \frac{\epsilon_1}{\epsilon_2}p_2$ and $p_1^2 - p_2$ (eigenfunctions of the trigonometric Calogero-Sutherland system) for partitions $(2,0)$ and $(1,1)$ respectively. It is easy to see that these Jack polynomials coincide with \eqref{es2} under isomorphism
\begin{equation}
\overline{a}_{-m} \vert 0 \rangle \longleftrightarrow -i \e_2 p_m\,. \label{map}
\end{equation}  

\subsubsection{Three soliton configuration}
A generic state with $k=3$ can be written as
\begin{equation}
(c_1 \overline{a}_{-1}^3 + c_2 \overline{a}_{-2}\overline{a}_{-1} + c_3 \overline{a}_{-3})\vert 0 \rangle
\end{equation}
The eigenvalue equation for $\widehat{I}_3$
\begin{equation}
\widehat{I}_3 (c_1 \overline{a}_{-1}^3 + c_2 \overline{a}_{-2}\overline{a}_{-1} + c_3 \overline{a}_{-3})\vert 0 \rangle
= E_3 (c_1 \overline{a}_{-1}^3 + c_2 \overline{a}_{-2}\overline{a}_{-1} + c_3 \overline{a}_{-3})\vert 0 \rangle
\end{equation}
leads to an equation for eigenvalue $E_3$ with three solutions

\begin{equation*}
\begin{split}
\frac{E_3^{(1)}}{\e_1 \e_2} \,=\,& i \dfrac{3}{2} (\e_1 + \e_2) + 3 i \e_1 + \tilde{p} \dfrac{3 i \e_2 (\e_1 + \e_2)}{2\e_1 - \e_2} \\
& - \tilde{p}^2 \dfrac{3 i \e_2 (22 \e_1^3 + 18 \e_1^2 \e_2 - 3 \e_1\e_2^2 + \e_2^3)}{(2\e_1 - \e_2)^3} + o(\tilde{p}^3)\,, \\ 
\frac{E_3^{(2)}}{\e_1 \e_2} \,=\,&  i \dfrac{5}{2} (\e_1 + \e_2) - \tilde{p} \dfrac{2 i (\e_1 + \e_2) (\e_1^2 - 7\e_1 \e_2 + \e_2^{2})}{2\e_1^2 - 5\e_1 \e_2 + 2 \e_2^{2}} \\
& + \tilde{p}^2 \dfrac{2 i (20\e_1^7 - 121\e_1^6 \e_2 + 6\e_1^5 \e_2^2 + 34\e_1^4 \e_2^3 + 34 \e_1^3 \e_2^4 + 6 \e_1^2 \e_2^5 - 121 \e_1 \e_2^6 + 20 \e_2^7)}{(2\e_1^2 - 5 \e_1 \e_2 + 2 \e_2^2)^3} + o(\tilde{p}^3)\,, 
\end{split}
\end{equation*}
\begin{equation}
\begin{split}
\frac{E_3^{(3)}}{\e_1 \e_2} \,=\,& i \dfrac{3}{2} (\e_1 + \e_2) + 3 i \e_2 + \tilde{p} \dfrac{3 i \e_1 (\e_1 + \e_2)}{2\e_2 - \e_1} \\
& - \tilde{p}^2 \dfrac{3 i \e_1 (22 \e_2^{3} + 18\e_2^2 \e_1 - 3 \e_2\e_1^{2} + \e_1^{3})}{(2\e_2 - \e_1)^3} + o(\tilde{p}^3)\,. \label{k3E3}
\end{split}
\end{equation}
Similarly, equation for $\widehat{I}_4$
\begin{equation}
\widehat{I}_4 (c_1 \overline{a}_{-1}^3 + c_2 \overline{a}_{-2}\overline{a}_{-1} + c_3 \overline{a}_{-3})\vert 0 \rangle
= E_4 (c_1 \overline{a}_{-1}^3 + c_2 \overline{a}_{-2}\overline{a}_{-1} + c_3 \overline{a}_{-3})\vert 0 \rangle
\end{equation}
only admits non-trivial solutions for the $E_4$ energies
\begin{equation}
\begin{split}
\frac{E_4^{(1)}}{\e_1 \e_2} = & -\left( \dfrac{\e_2^2}{2} + \dfrac{9 \e_1 \e_2}{4} + \dfrac{9\e_1^2}{2} \right) - \tilde{p} \dfrac{3 \e_2 (\e_1 + \e_2) (5 \e_1 + \e_2)}{2(2 \e_1 - \e_2)} \\
& + \tilde{p}^2 \dfrac{3 \e_2 (\e_1 + \e_2) (47 \e_1^3 + 2 \e_1^2 \e_2 + \e_1 \e_2^2 + \e_2^3)}{(2\e_1 - \e_2)^3} + o(\tilde{p}^3)\,, \\ 
\frac{E_4^{(2)}}{\e_1 \e_2} = & -\left( \dfrac{3\e_1^{2}}{2} + \dfrac{7 \e_1 \e_2}{4} + \dfrac{3\e_2^2}{2} \right) + \tilde{p} \dfrac{(\e_1 + \e_2)^2 (\e_1^2 - 13 \e_1 \e_2 + \e_2^{2})}{2\e_1^2 - 5\e_1 \e_2 + 2 \e_2^{2}} \\
& - \tilde{p}^2 \dfrac{(\e_1 + \e_2)^2 (40 \e_1^6 - 303 \e_1^5 \e_2 + 345 \e_1^4 \e_2^2 - 325 \e_1^3 \e_2^3 + 345 \e_1^2 \e_2^4 - 303 \e_1 \e_2^5 + 40 \e_2^6)}{(2\e_1^2 - 5 \e_1 \e_2 + 2 \e_2^{2})^3} + o(\tilde{p}^3)\,, \\  
\frac{E_4^{(3)}}{\e_1 \e_2} = & -\left( \dfrac{\e_1^{2}}{2} + \dfrac{9 \e_1 \e_2}{4} + \dfrac{9\e_2^{2}}{2} \right) - \tilde{p} \dfrac{3 \e_1 (\e_1 + \e_2) (5 \e_2 + \e_1)}{2(2 \e_2 - \e_1)} \\
& + \tilde{p}^2 \dfrac{3 \e_1 (\e_1 + \e_2) (47 \e_2^{3} + 2 \e_2^2 \e_1 + \e_2 \e_1^2 + \e_1^3)}{(2\e_2 - \e_1)^3} + o(\tilde{p}^3)\,. \label{k3E4}
\end{split}
\end{equation}
One can check that the $\gamma$ expansions at orders $\gamma^3$, $\gamma^4$ of \eqref{300osc}, \eqref{210osc} and \eqref{111osc} reproduce \eqref{k3E3} and \eqref{k3E4}. Again, in the limit $\tilde{p} \rightarrow 0$ the eigenstates are mapped to Jack polynomials under \eqref{map}.
\\

\bibliography{cpn1}
\bibliographystyle{JHEP}

\end{document}